\documentclass[a4paper,11pt]{article}

\pdfoutput=1 % if your are submitting a pdflatex (i.e. if you have
\usepackage{jheppub} % for details on the use of the package, please
\usepackage{hyperref}
\usepackage{makeidx}
\usepackage[totoc]{idxlayout}
\usepackage[T1]{fontenc} % if needed
\definecolor{darkgreen}{rgb}{0.0, 0.5, 0.0}
\usepackage{amsmath}
\usepackage{hhline}
\usepackage{multirow}
\usepackage{rotating}
\usepackage{mathdots}
\usepackage{slashed}

\usepackage{cancel}

\usepackage{slashed}
\usepackage{url}

\usepackage{amssymb}  % more math symbols
\usepackage{float} % I think this allows fixation of graphics to certain locations using [H!]-like commands
\usepackage{xcolor} % see color package
\usepackage{graphicx} % takes care of graphic including machinery
\usepackage{mathdots} % for dots in the matrices
\usepackage[normalem]{ulem}
\usepackage{bbm}
\usepackage{dsfont}
\usepackage[mathscr]{euscript}
\usepackage{ntheorem}
\usepackage{tensor} % Allows for the \tensor command which improves typsetting of indices
\usepackage{pdfpages} % Allows inclusion of pdf files
\usepackage[shortlabels]{enumitem} % Allows changes of enumerate counter

\newcommand{\bB}{\blue{\bar{B}}}

\newcommand{\phij}{\blue{\phi^j}}

\newcommand{\KS}{{\blue{\mathrm{KS}}}}

\newcommand{\dbmuisq}{\red{(d\bmu^i)^2}}
\newcommand{\bmuisq}{\red{(\bmu^i)^2}}
\newcommand{\muisq}{\red{(\mu^i)^2}}
\newcommand{\phii}{\red{\phi^i}}

\newcommand{\Hor}{{\blue{{\mathbf{H}}}}}

\newcommand{\btau}{\blue{\tau}}

{\catcode `\@=11 \global\let\AddToReset=\@addtoreset}
\AddToReset{equation}{section}

{\catcode `\@=11 \global\let\AddToReset=\@addtoreset}
\AddToReset{figure}{section}
\renewcommand{\thefigure}{\thesection.\arabic{figure}}

{\catcode `\@=11 \global\let\AddToReset=\@addtoreset}
\AddToReset{table}{section}
\renewcommand{\thetable}{\thesection.\arabic{table}}

\newcommand{\mcU}{{\mycal U}}

\newcommand{\fourg}{\blue{\mathbf{g}}}

\newcommand{\myGauss}{{\blue{\twoscsign}}}

\newcommand{\twoscsign}{{\blue{\varepsilon}}}

\newcommand{\red}[1]{{\color{red} #1}}

\newcommand{\R}{\mathbb{R}}

\newcommand{\zhTBW}{\blue{\gamma}}
\newcommand{\zzhTBW}{\blue{\mathring{\zhTBW}}}

\newcommand{\mcM}{{\mycal M}}
\newcommand{\mcN}{{\mycal N}}
\newcommand{\scri}{{\mycal I}}%
\newcommand{\Scri}{\scri}

\newcommand{\ringh}{{  \zzhTBW }}%

\newcommand{\eeal}[1]{\label{#1}\end{eqnarray}}

\newtheorem{theorem}{\sc  Theorem\rm}[section]

\newtheorem{lemma}[theorem]{\sc Lemma\rm}
\newtheorem{Lemma}[theorem]{\sc Lemma\rm}

\newtheorem{Proposition}[theorem]{\sc Proposition\rm}

\theorembodyfont{}

\DeclareFontFamily{OT1}{rsfs}{}
\DeclareFontShape{OT1}{rsfs}{m}{n}{ <-7> rsfs5 <7-10> rsfs7 <10-> rsfs10}{}
\DeclareMathAlphabet{\mycal}{OT1}{rsfs}{m}{n}

\newcommand{\mnote}[1]
{\protect{\stepcounter{mnotecount}}$^{\mbox{\footnotesize $%\!\!\!\!\!\!\,
		\bullet$\themnotecount}}$ \marginpar{%\color{red}%
	\raggedright\tiny\em $\!\!\!\!\!\!\,\bullet$\themnotecount: #1} }

\newcommand{\zgamma}{\blue{\mathring \gamma}}

\newcommand{\Z}{\mathbbm{Z}}

\newcommand{\blue}[1]{{\color{blue} #1}}

\newcommand{\mysigma}{{{\sigma}}}

\newcounter{mnotecount}[section]

\renewcommand{\themnotecount}{\thesection.\arabic{mnotecount}}

\newcommand{\mnotex}[1]%{}
{\protect{\stepcounter{mnotecount}}$^{\mbox{\footnotesize
		$%\!\!\!\!\!\!\,
		\bullet$\themnotecount}}$ \marginpar{%\color{red}%
	\raggedright\tiny\em
	$\!\!\!\!\!\!\,\bullet$\themnotecount: #1} }

\newcommand{\ptcno}[1]{}

\newcommand{\barg}{\blue{\bar g}}

\newcommand{\eqrefl}[1]{\eqref{#1}}

\newcommand{\bel}[1]{\begin{equation}\label{#1}}
\newcommand{\bea}{\begin{eqnarray}}
\newcommand{\bean}{\begin{eqnarray}\nonumber}
\newcommand{\beal}[1]{\begin{eqnarray}\label{#1}}
\newcommand{\eea}{\end{eqnarray}}

\newcommand{\nn}{\nonumber}

\newcommand{\qed}{\hfill $\Box$}
\newcommand{\qedskip}{\hfill $\Box$\medskip}
\newcommand{\proof}{\noindent {\sc Proof:\ }}
\newcommand{\be}{\begin{equation}}
\newcommand{\eeq}{\end{equation}}
\newcommand{\ee}{\end{equation}}
\newcommand{\beqa}{\begin{eqnarray}}
\newcommand{\eeqa}{\end{eqnarray}}
\newcommand{\beqan}{\begin{eqnarray*}}
\newcommand{\eeqan}{\end{eqnarray*}}
\newcommand{\ba}{\begin{array}}
\newcommand{\ea}{\end{array}}

\newcommand{\ptcheck}[1]{{\color{darkgreen}\mnotex{ptc : checked on #1}}}
\newcommand{\wancheck}[1]{{\color{darkgreen}\mnotex{wan : checked on #1}}}

\def\beq{\begin{eqnarray}}
\def\eeq{\end{eqnarray}}

\def \C{{\mathbb{C}}}

\def\be{\begin{equation}}
\def\ee{\end{equation}}
\def\bea{\begin{eqnarray}}
\def\eea{\end{eqnarray}}

\newcommand{\mcH}{\mycal H}

\newcommand{\newg}{\blue{\breve{g}}}

\newcommand{\nobara}{\blue{a}}

\newcommand{\myPhi}{\blue{H}}
\newcommand{\myHS}{\blue{\myH|_{S^{N-1}}}}

\newcommand{\bnabla}{\blue{\bar\nabla}}

\newcommand{\palphasquare}{\ellsquare}
\newcommand{\ellsquare}{\red{\ell^{-2}} }

\newcommand{\barVh}{\red{\bar V}}

\newcommand{\mcB}{\blue{\mycal B}} 

\newcommand{\bgamma}{\blue{\breve{\gamma}}}

\newcommand{\brU}{\blue{\breve{U}}}

\newcommand{\bg}{\blue{\bar{g}}}

\newcommand{\myH}{\blue{\bar{H}}}
\newcommand{\myF}{\blue{\bar{F}}}

\newcommand{\barU}{\red{\br^{-1}  \barU}}
\newcommand{\barUh}{\red{  \bar U}}

\newcommand{\bUh}{\barUh}

\newcommand{\bmu}{\bar{\mu}}
\newcommand{\bt}{\bar{t}}
\newcommand{\br}{\bar{r}}
\newcommand{\baa}{\bar{a}}
\newcommand{\bm}{\bar{m}}
\newcommand{\bW}{\bar{W}}

\newcommand{\bXi}{\bar{\Xi}}

\newcommand{\N}{\mathbb{N}}

\newcommand{\FGn}[1]{{\mnote{{\bf finn:}
	#1 }}}

\title{\boldmath Kerr-AdS type higher dimensional black holes with non-spherical cross-sections of horizons%
\protect
\footnote{Preprint:UWThPh-2024-24
}
}

 \author{Piotr T.\ Chru\'sciel,}
 \author{Wan Cong,}
 \author[2]{and  Finnian Gray\note{Corresponding author}} \affiliation{University of Vienna, Faculty of Physics
  \\Boltzmanngasse 5, A 1090 Vienna, Austria}

\emailAdd{piotr.chrusciel@univie.ac.at}
\emailAdd{wan.cong@univie.ac.at}
\emailAdd{finnian.gray@univie.ac.at}

\abstract{We  construct, in even spacetime dimensions, a family of  singularity-free Kerr-anti-de Sitter-like  black holes with negatively curved cross-sections of  conformal infinity and  { 
 non-compact, non-spherical cross-sections of horizons, with the full set of rotation parameters. 
 }
}

\renewcommand{\red}[1]{#1}
\renewcommand{\blue}[1]{#1}
\renewcommand{\wancheck}[1]{}
\renewcommand{\ptcheck}[1]{}
\renewcommand{\FGn}[1]{}

\begin{document}
\maketitle

\flushbottom

\section{Introduction}

The Schwarzschild metrics, together with their higher-dimensional generalisations, are expected to exhaust  the collection of  $(n+1)$-dimensional, vacuum, static, asymptotically flat metrics satisfying natural global properties~\cite{ChCo}.%  
\footnote{See~\cite[Remark~1.4 in v2]{ChodoshEtAl}  and references therein concerning the positive energy theorem; that last theorem is  used  
	in the proof of uniqueness.} 
These metrics are a special case of the Birmingham-Kottler metrics~\cite{Birmingham,Kottler}, which allow for any value of the cosmological constant $\Lambda$, and where the usual $(n-1)$-dimensional spherical sections of the event horizon can be replaced by any $(n-1)$-dimensional Einstein geometry.

The Kerr metric is the unique $(3+1)$-dimensional vacuum stationary analytic and asymptotically flat black hole satisfying a set of reasonable properties, and  its higher dimensional generalisation are likewise of interest. 
The family of higher dimensional, asymptotically flat Kerr-like vacuum metrics was introduced by Myers and Perry in~\cite{Myers:1986un} while the case in which we are interested, with non-zero cosmological constant, has been presented in~\cite{Gibbons:2004uw,Gibbons:2004js}. 
In $(3+1)$ dimensions the algebraically special property of the Weyl tensor   of the Kerr metric is often emphasised, and such algebraically special metrics have been considered and classified in higher dimensions, see~\cite{Ortaggio:2017ydo,Reall:2011ys} and references therein; some partial uniqueness results have been obtained~\cite{Lucietti:2020ltw,Hollands:2012xy}, see  also \cite{MarsKdS} for a uniqueness result when $\Lambda>0$.
The Kerr-like class of metrics has  been generalised to include multiple NUT parameters~\cite{Chen:2006ea}. 
We refer the interested reader to the review~\cite{Emparan:2008eg} for more examples of higher dimensional black holes.

Now, the sections of the horizons of the metrics of~\cite{Gibbons:2004uw,Gibbons:2004js} have spherical topology.  
{
	%\blueref{points 1 and 2, to acknowledge the work of Klemm et al and clarify the topology of horizon} 
	A more general class of Kerr-AdS-like metrics, with negatively curved horizons and 
	with a single rotation parameter has been discovered by Klemm, Moretti, and Vanzo~\cite{Klemm:1997ea,Klemm:1998kd}. 
	This begs question of existence of Kerr-like $(n+1)$-dimensional vacuum metrics with negatively curved horizons and with the whole set of rotation parameters. 
}
The aim of this paper is to point out the existence of such stationary  metrics, in the case of a negative cosmological constant, with non-spherical sections of the horizon, in all even spacetime dimensions $(n+1) \ge 4$. 

The metrics we obtain are parameterised by a mass parameter $\bm$ and by $N:=(n-1)/2$ rotation parameters $\baa_i$.  
As already pointed out, they generalise the metrics presented in~\cite{Klemm:1997ea,Klemm:1998kd}, which have only one rotation parameter.
{
	%\blueref{added discussion about geodesic completeness and global hyperbolicity}
	In particular, when $\bar m \bar{a}_1...\bar{a}_N\neq 0$ the metrics have the remarkable property of having no curvature singularities. 
	We have not investigated the question, whether or not the spacetimes are geodesically complete. 
	It should be clear from the projection diagrams that we  construct in Section~\ref{s15XI24.1} that the maximal extensions considered there are not globally hyperbolic; whether or not the spacetimes are geodesically complete, and whether the domains of outer communication are globally hyperbolic, would require further investigations.
} 
Again remarkably, no causality violations arise when
\begin{equation}\label{25XI24.31}
2 |\bm| <  \prod_{i=1}^N |\baa_i|^{\frac{2N-1}{N}} 
\,.
\end{equation}
(This inequality provides a sufficient condition; an if-and-only-if condition can be found in Proposition~\ref{P24XI24.1} below.) 
Note that the inequality \eqref{25XI24.31} goes against  the intuition gained from the $(3+1)$-dimensional Kerr metric, where large angular momenta lead to singularities.

It is natural to enquire about global Hamiltonian charges associated with our metric. 
Such charges are usually defined by  integration of Noether-type integrands over a sequence of boundaries which recede to infinity, and passing to the limit. However, we show that there are no natural compact such boundaries unless all rotation parameters vanish, which results in infinite integrals before even passing to infinity. 
This appears to considerably reduce the interest of our solutions. 

We note that the regularity of the metric does not enforce a lower bound on the parameter $\bm$ appearing in our solution, which is usually considered to be directly proportional to the mass when the rotation parameters vanish. 
This lack of lower bound might well be a feature related to the non-compactness above. 

The metrics have a timelike conformal infinity \emph{\`a la Penrose}. 
The family includes metrics with Killing horizons with vanishing surface gravity $\kappa$, with the outermost Killing horizon having non-zero $\kappa$, as seen from the asymptotic region where $\bar m \bar r \to \infty$.

This work is organised as follows: In Section~\ref{Kerr-dS appendix} we recall the higher-dimensional Kerr-type metrics with a cosmological constant. 
In Section~\ref{s24XI24.1} we present our new metrics. 
We show in Section~\ref{ss24XI24.3} that the metrics reduce to the Birmingham-Kottler metrics when the rotation parameters vanish. 
In Sections~\ref{sec: singularities}-\ref{17X24f.1} we establish extendibility across Killing horizons by rewriting the metric in a Kerr--Schild form, and show nonexistence of ``non-coordinate'' singularities. 
In Section~\ref{ss28VII24.4} we construct a conformal completion at infinity. The signature of the metric is checked in Section~\ref{ss28VII24.1}.
We give a sharp criterion on the parameters for absence of causality violations in Section~\ref{ss24XI24.1}, see Proposition~\ref{P24XI24.1}
there. 
We show that the geometry of sections of the horizon cannot be realised on a compact manifold for small rotation parameters in Section~\ref{ss28VII24.2}; whether or not this can be done in general remains to be seen. Existence of black hole regions is established in Section~\ref{s14XI24.1}. 
In Section~\ref{s15XI24.1} we construct projection diagrams for a class of natural maximal analytic extensions. 
The question of generalisation of the construction to other dimensions, positive cosmological constant, or Ricci flat sections of the horizon is briefly addressed in Section~\ref{App24XI24.1}. 
Some technical results, in particular the analysis of the number of zeros of the equation determining the location  of Killing horizons, are left to Appendices~\ref{s5XI24.1}-\ref{s17XI24.1}.

\section{Spherical Kerr-Anti-de Sitter black holes in higher dimensions}
\label{Kerr-dS appendix}

We start by recalling the explicit form of  the Kerr-AdS metrics% of Gibbons et al.
~\cite{Gibbons:2004js,Gibbons:2004uw}, which generalise the multiply rotating Myers-Perry black holes~\cite{Myers:1986un} to include a cosmological constant. 

The Kerr-AdS metric in $d=(n+1)$-spacetime dimensions with $\Lambda <0$, is given by\cite{Gibbons:2004js,Gibbons:2004uw}:%
\footnote{We  note that our $a_i$'s are the negatives of the ones in~\cite{Gibbons:2004js}.}
\begin{align}\label{eq:KerrAdS}
ds^2=g_{\mu\nu}dx^\mu dx^\nu
=&-W(1+\ellsquare r^2)dt^2+\frac{2m}{U}\Bigl(Wdt+\sum_{i=1}^\red{N} \frac{a_i \muisq d\red{\phi^i}}{\Xi_i}\Bigr)^2\nonumber
\\
&+\sum_{i=1}^\red{N}\frac{r^2+a_i^2}{\Xi_i}\bigl(\muisq  \red{(d\phi^i)^2}+\red{(d\mu^i)^2})
+\frac{Udr^2}{V-2m}+\delta r^2 \red{(d\mu^{\red{N}+\delta})^2}
\nonumber\\
&-\frac{1}{\ell^2 W(1+\ellsquare r^2)}\Bigl(\sum_{i=1}^\red{N} \frac{r^2+a_i^2}{\Xi_i}\red{\mu^i} d\red{\mu^i}+\delta r^2 \red{\mu^{\red{N}+\delta}} d\red{\mu^{\red{N}+\delta}}\Bigr)^2\,,
\end{align}
with $\ell^2>0$, where  
\begin{align}\label{eq: Metfuncs}
W&=\sum_{i=1}^{\red{N}} \frac{\muisq}{\Xi_i}+\delta (\red{\mu^{\red{N}+\delta}})^2\,,
\quad
V=r^{\delta-2}(1+\ellsquare r^2)\prod_{i=1}^\red{N}(r^2+a_i^2)\,,\nonumber\\
U&%=\frac{V}{1+\ellsquare r^2}\Bigl(1-\sum_{i=1}^\red{N}\frac{a_i^2\muisq}{r^2+a_i^2}\Bigr)
=r^\delta\left( \sum_{i=1}^{N+\delta}\frac{\muisq}{r^2+a_i^2}\right)\prod_{i=1}^\red{N}(r^2+a_i^2)\,,
\quad 
\Xi_i=1-\ellsquare  a_i^2
\,.
\end{align}
Here, $\delta=1, 0$ for even, respectively, odd spacetime dimensions, $N=\bigl[\frac{n}{2}\bigr]$ (where $[A]$ denotes the integer part of $A$) so that $d=n+1=2N+\delta+1$, with $a_{N+\delta}=0$ when $\delta=1$. The coordinates $\red{\mu^i}$ obey the constraint
\begin{equation}\label{3VI24.f1}
\sum_{i=1}^{N+\delta} \muisq=1\,.
\end{equation}
In these coordinates 
the angles $\red{\phi^i}$ are periodic with period $2\pi$. 
The metrics satisfy the vacuum Einstein equations,
$G_{\mu\nu}+\Lambda g_{\mu\nu}=0$, with cosmological constant $\Lambda=-\frac 1 2n(n-1)\ellsquare  $, see e.g.~\cite{Hamamoto:2006zf} where the curvature is calculated explicitly in all dimensions%
\footnote{A translation between the coordinates systems used here and those of  Myers--Perry can be found in Sections 4.2-5 of~\cite{Frolov:2017kze}.}%
.

There are many questions concerning this geometry which do not seem to have been satisfactorily addressed in the literature: are the domains of  outer communication  globally hyperbolic? or at least stably causal? when are singularities and causality violations hidden beyond event horizons? 
See however the review~\cite{Emparan:2008eg} in general, \cite{Myers:1986un,Myers:2011yc} for some discussion of these issues with vanishing cosmological constant, and  \cite{Gibbons:2004js,Gibbons:2004uw,Frolov:2017kze} for a brief discussion with $\Lambda\ne0$.

 \section{Multiply rotating  Kerr-AdS-type black holes  with non-spherical  horizons}
 \label{s24XI24.1} 
 
 From now on, \emph{unless explicitly indicated otherwise} we consider the metrics \eqref{eq:KerrAdS} with  (i) all rotation parameters nonzero, and (ii) even $d$, thus $\delta = 1$ and $N = \frac{d-2}{2}$.

 To construct our new family of metrics, we will perform an analytic continuation of the metric \eqref{eq:KerrAdS}, similar to that done in~\cite{Klemm:1997ea,Klemm:1998kd}. 
 First, we redefine the constrained variables according to
 \begin{align}
 \mu_j&= i \bmu_j \,,\quad 1\leq j<N\,,\qquad \text{and}\qquad
 \mu^{N+1}=\red{\bmu^{N+1}}
 \,,
 \end{align}
 with $\bar \mu_j\in\R$, where $i =\sqrt{-1}$.  
 Then the constraint \eqref{3VI24.f1} becomes
 \begin{equation}\label{3VI24.f5}
 \red{(\bmu^{N + 1})^2}  =	1 +   \sum_{i=1}^{N} \bmuisq
 \,,
 \end{equation}
 which resembles how one would embed an $N$-dimensional hyperbolic space in one higher dimension. 
 To retain the signature $(-,+,\ldots,+)$ of the metric, we  make the further substitutions
 \begin{equation}\label{3VI24.f6}
 t=i \bar{t}\,,\quad r= -i \bar{r}\,, \quad \,\quad a_j=-i \baa_j\,,\quad m= {(-1)^{N+1}i \bar{m} }\,,
 \end{equation}
 where all of $\bar t$, $\bar r$, $\bar a_j$ and $\bar m$ are taken to be real valued.

 We define metric functions analogous to \eqref{eq: Metfuncs} as
 \begin{align}\label{24VII24.f1}
 \bar{W}&: = \red{(\bmu^{N+1})^2}  -\sum_{i=1}^{N} \frac{\bmuisq}{\bar{\Xi}_i}
 \,,
 \quad 
 \barVh=
 -(1-\ellsquare \bar{r}^2)\prod_{i=1}^N(\bar{r}^2+\baa_i^2)\,,\nonumber
 \\
 \barUh&
 : = \left( {\red{(\bmu^{N+1})^2}}  -{\bar{r}^2} \sum_{i=1}^{N }\frac{\bmuisq}{\bar{r}^2+\baa_i^2}\right)\prod_{i=1}^N(\bar{r}^2+\baa_i^2)\,,
 \quad
 \bar{\Xi}_i :=1 +\ellsquare  \baa_i^2   
 \,.
 \end{align}

 With the above redefinitions the metric \eqref{eq:KerrAdS} takes the form
 \begin{align}\label{3VI24.f7}
 \bar g =\bar{g}_{\mu\nu}dx^\mu dx^\nu=&\bar{W}(1-\ellsquare\bar{r}^2)d\bar{t}^2
 +
 \frac{\barUh d\bar{r}^2}{\barVh-2\bar{m}\br}
 +
 \frac{2\bar{m}\br}{\barUh}
 \biggl(\bar{W}d\bar{t}  +  \sum_{i=1}^N \frac{\baa_i \bmuisq d\red{\phi^i}}{\bar{\Xi}_i}   
 \biggr)^2
 \nonumber
 \\
 &
 +\sum_{i=1}^N\frac{\bar{r}^2+\baa_i^2}{\bar{\Xi}_i}\bigl(\bmuisq\red{(d\phi^i)^2}+\dbmuisq )
 - \bar{r}^2 \red{(d\red{\bmu^{N+1}})^2}
 \nonumber
 \\
 &-
 \frac{ 1}{\bar{W}(\ell^2 -  \bar{r}^2)}
 \Bigl(\sum_{i=1}^N \frac{\bar{r}^2+\baa_i^2}{\bar{\Xi}_i}\red{\bmu^i} d\red{\bmu^i}
 - \br^2 \red{\bmu^{N+1}}d\red{\bmu^{N+1}}\Bigr)^2\,.
 \end{align}

 Choose a mass parameter $\bar m\in \R$ and a cosmological constant parameter $\ell>0$. 
 Consider any connected component, say $\mcU$,  of the set on which $\bar g$ is defined and non-degenerate. 
 Proposition~\ref{P6XI24.1} below shows that such sets do not depend upon the rotation parameters, hence all the metrics $\bar g$ with the same value of $\bar m$ are simultaneously defined there. 
 The calculations in~\cite{Hamamoto:2006zf} 
 {
 	%\blueref{Clarified the role of the calculations in \cite{Hamamoto:2006zf} and added an appendix for the analytic continuation arguments.} 
 	(which take place in a different coordinate system to those used here; a translation between the systems can be found in Sections 4.2-5 of~\cite{Frolov:2017kze}) demonstrate that \eqref{eq:KerrAdS} satisfies the Einstein equations in all dimensions.
 	This, together with the usual analytic continuation arguments (see Appendix \ref{s20V25.1} where they are spelled-out in detail), shows that the tensor field $\bar g$} defines  on $\mcU$ a pseudo-Riemannian metric which solves the vacuum Einstein equations with cosmological constant 
 \begin{equation}\label{4XI24.2}
 \Lambda = -\frac 1 2n(n-1)\ellsquare
 \,.
 \end{equation}
 Note that analytic continuation could change the signature, but we show that this does not happen in our case in Section~\ref{ss28VII24.1}.

 Using the constraint \eqref{3VI24.f5} we can eliminate $\bmu^{N+1}$ to obtain  
 \begin{align}
 \label{22X24.1}
 &
 \bar W=1 +  \sum_{i=1}^{N }
 \frac{ \bar a_i^2\bmuisq}{\ell^2 \bar\Xi_{i}}
 \ge 1
 \,,
 \qquad 
 \bUh    =  \bigg(1
 + \sum_{i=1}^N \frac{ {\bmuisq}\baa_i^2}
 {\br^2+\baa_i^2}\bigg) \prod_{i=1}^N (\bar{r}^2+\baa_i^2)
 \ge \prod_{i=1}^N  \baa_i^2 
 \,,
 \\
 &
 \barVh=
 (\ellsquare \bar{r}^2-1)\prod_{i=1}^N(\bar{r}^2+\baa_i^2)\,,
 \qquad
 \bar{\Xi}_i =1 +\ellsquare  \baa_i^2  \ge 1
 \,. 
 \end{align}

 Next, using again  \eqref{3VI24.f5} we find 
 \begin{align} 
 \sum_{i=1}^N \frac{\bar{r}^2+\baa_i^2}{\bar{\Xi}_i}\red{\bmu^i} d\red{\bmu^i}
 - \br^2 \red{\bmu^{N+1}}d\red{\bmu^{N+1}}
 =
 (1-\ellsquare \bar{r}^2)\sum_{i=1}^N \frac{\baa_i^2}{\bar{\Xi}_i}\red{\bmu^i} d\red{\bmu^i} 
 =  \frac{ \ell^2- \bar{r}^2}{2} d \bar W
 \,.
 \end{align}
 This allows us to rewrite the metric as   
 \begin{align}\label{3VI24.f78}
 \bar g  =&
 \bar{W}(1-\ellsquare\bar{r}^2)d\bar{t}^2
 +\frac{\barUh d\bar{r}^2}{\barVh-2\bar{m}\br}
 +\frac{2\bar{m}\br}{\barUh}
 \biggl(\bar{W}d\bar{t} 
 +  \sum_{i=1}^N \frac{\baa_i \bmuisq d\red{\phi^i}}{\bar{\Xi}_i}   
 \biggr)^2
 \nonumber
 \\
 & 
 +\sum_{i=1}^N\frac{\bar{r}^2+\baa_i^2}{\bar{\Xi}_i}\bigl(\bmuisq\red{(d\phi^i)^2}+\dbmuisq ) 
 - \bar{r}^2 
 \frac {\Big( \tfrac{1}{2}\sum_{i=1}^{N} d(\bmuisq )\Big)^2}{1 +   \sum_{i=1}^{N} \bmuisq}  
 \nonumber
 \\
 &
 %+
 -\frac{ \ell^2- \bar{r}^2}{4 \bar W}
 ( d\bar W)^2\,,
 \end{align}
 where we didn't expand the differentials $d(\bmuisq )$ to make it clear that they define smooth covectors fields on the planes parameterised by the polar coordinates $(\red{\bmu^i},  \red{\phi^i})$. 
 Indeed, the change of coordinates  
 \begin{equation}
 (\red{x^i},\red{y^i} )=\big (\red{\bmu^i}\cos(\red{\phi^i}),\red{\bmu^i}\sin(\red{\phi^i})\big)
 \label{20XI24.1}
 \end{equation}
 leads to
 $$
 d(\bmuisq ) = 2 ( \red{x^i} d\red{x^i} +\red{y^i}  d\red{y^i} )
 \,,
 \quad
 \bmuisq d\red{\phi^i} = \red{x^i} d\red{y^i}  - \red{y^i}  d\red{x^i}
 \,,
 \quad
 \bmuisq\red{(d\phi^i)^2}+\dbmuisq  = (d\red{x^i})^2 + (d\red{y^i})^2
 \,,
 $$
 which renders $\bar g$ manifestly smooth at the axes of rotation. Note also that all remaining functions appearing in the metric are smooth functions of $\muisq= (\red{x^i})^2+(\red{y^i})^2$, hence smooth with respect to the differentiable structure defined by the $(\red{x^i},\red{y^i} )$'s.
 
 We now proceed to analyse the global structure of the family of metrics given by \eqref{3VI24.f78}.

 \subsection{Non-rotating case}
 \label{ss24XI24.3}
 
 When all the $\baa_i$'s vanish, using the constraint \eqref{3VI24.f5} and $d=n+1=2N+2$, we have 
 \begin{align}
 \bar{W}&=1%\red{-1}\sum_{i=1}^N \frac{\bmuisq}{\bar{\Xi}_i \,\sigma^{-2}}+ \frac{\red{(\bmu^{N+1})^2}}{\bar{\Xi}_{N+1}\,\sigma^{-2}}
 \,,
 \quad \barVh= - \bar{r}^{n-1}(1 -\ellsquare \bar{r}^2)
 %\bar{r}^{1-2}(\red{-1}+\ellsquare \bar{r}^2)\prod_{i=1}^N(\bar{r}^2 )
 \,,%\nonumber\\
 \quad 
 \barUh=\bar{r}^{n-1}
 %=\bar{r}^1\left(\frac{\red{(\bmu^{N+1})^2}}{\bar{r}^2 }+ \red{-1}\sum_{i=1}^{N+1-1}\frac{\bmuisq}{\bar{r}^2 }\right)\prod_{i=1}^N(\bar{r}^2 )
 \,,
 \quad
 \bar{\Xi}_i=1 \,,
 \end{align}
 which simplifies the metric to
 %P
 \begin{align}\label{3VI24.f7asdf}
 \barg 
 = &  
 \
 \bar{W}(1-\ellsquare \bar{r}^2)d\bar{t}^2+\frac{\barUh d\bar{r}^2}{\barVh-2\bar{m}\br}
 +\frac{2\bar{m}\br}{\barUh}\big(\bar{W}d\bar{t}
 \big)^2
 % \nonumber\\
 % &
 +\sum_{i=1}^N\frac{\bar{r}^2 }{\bar{\Xi}_i}\bigl(\bmuisq\red{(d\phi^i)^2}+\dbmuisq )
 \nonumber
 \\
 &
 -\br^2 d\red{\bmu^{N+1}}
 -\frac{\ellsquare }{\bar{W}(1 -\ellsquare \bar{r}^2)}
 \Bigl(
 \underbrace{
 	\sum_{i=1}^{{N}} \frac{\bar{r}^2 }{\bar{\Xi}_i}\red{\bmu^i} d\red{\bmu^i}
 	- \br^2\red{\bmu^{N+1}}(d\red{\bmu^{N+1}})^2
 }_{=r^2d(\sum_i\bmuisq-\red{(\bmu^{N+1})^2})=0}
 \Bigr)^2\,
 \nonumber
 \\
 = & \ (1 -\ellsquare \bar{r}^2  + 2\bar{m}\bar{r}^{n-2} )d\bar{t}^2
 -\frac{d\bar{r}^2}{(1 -\ellsquare \bar{r}^2 + 2\bar{m}\bar{r}^{n-2} )}
 \nonumber
 \\
 &
 +\bar{r}^2
 \underbrace{
 	\left(
 	\sum_{i=1}^{{N}} (\dbmuisq + \bmuisq \red{(d\phi^i)^2})
 	-\red{(d\red{\bmu^{N+1}})^2} \right)
 }_{\gamma_{AB}d\bar{x}^Ad\bar{x}^B}
 \nonumber
 \\
 = & \,
 -f(\bar{r})d\bar{t}^2+\frac{d\bar{r}^2}{f(\bar{r})}+\bar{r}^2\gamma_{AB}d\bar{x}^Ad\bar{x}^B\,,
 \end{align}
 where
 \begin{equation}\label{1VIII24.2}
 f(r)= -1 +\ellsquare \bar{r}^2 - \frac {2\bar m} {\bar{r}^{2N-1}}
 %1 -\ellsquare \bar{r}^2 - \frac {2\bar m} {\bar{r}^{2N+1-2}} 
 \,.
 \end{equation}
 Note that the last term after the first equality in \eqref{3VI24.f7asdf} vanishes because it is proportional to the exterior derivative of the constraint \eqref{3VI24.f5}.

 We now check for completeness that the metric on the transverse space is the expected negatively curved one.
 As before let $(\red{x^i},\red{y^i} )= (\red{\bmu^i} \cos \red{\phi^i},\red{\bmu^i} \sin \red{\phi^i})$. 
 Ignoring the constraint 
 \begin{equation}\label{3VI24.f5wer}
 (\bmu^{N+1})^2 =	1 +\sum_{i=1}^{N} \bmuisq
 \,,
 \end{equation}
 the metric $\gamma_{AB}d\bar{x}^Ad\bar{x}^B$ can be rewritten as
 \begin{align}
 \gamma_{AB}d\bar{x}^Ad\bar{x}^B
 &   
 =
 \sum_{i=1}^{{N}} 
 (\dbmuisq + \bmuisq \red{(d\phi^i)^2})
 -
 \red{(d\red{\bmu^{N+1}})^2}
 \nonumber
 \\
 & 
 =
 \sum_{i=1}^{N} \left((d\red{x^i})^2 + (d\red{y^i})^2\right)
 -
 \red{(d\red{\bmu^{N+1}})^2}  \,.
 \label{1VIII24.1}
 \end{align}
 This is the Minkowski metric in $2N+1$ dimensions, which induces the hyperbolic metric on the spacelike hyperboloid \eqref{3VI24.f5wer}.

 Finally, we can calculate the determinant directly when $\baa_i=0$. This gives,
 \begin{align}
 \det (\bar g_{\mu\nu})&=-\frac{\br^{4}}{\red{(\bmu^{N+1})^2}}\left(\frac{\red{(\bmu^{N+1})^2}}{\bar{r}^2} - \sum_{i=1}^{N}\frac{\bmuisq}{\bar{r}^2}\right)^2\prod_{i=1}^{N} \bar{r}^4\,\bmuisq
 \nonumber\\
 &=-\frac{\bar{r}^{2(d-2)}}{\bar\mu_{N+1}^{2}} \prod_{i=1}^{N} \bmuisq \,,
 \end{align}
 and provides a crosscheck of~\eqref{26VII24f.1}.
 
 Happily, we have recovered the Birmingham--Kottler metric~\cite{Birmingham,Kottler}, 
 {
 	%\redref{negative sign added in front of $d\bar t^2$ term}
 	%
 	\begin{equation}
 	\label{21X24/1}
 	g_{\mathrm{BK}}= -f_{\myGauss}d\bar{t}^2+\frac{d\bar{r}^2}{f_{\myGauss}} +\bar{r}^2\ringh_{AB}dx^Adx^B\,,\quad f_{\myGauss}=\myGauss+\ellsquare \bar{r}^2-2m\bar{r}^{n-2}\,.
 	\end{equation}
 }
 with $\myGauss=-1$, in Boyer--Lindquist coordinates, with $\palphasquare  = -\frac{2\Lambda}{n(n-1)}$: 
 \begin{equation}
 \barg 
 \big|_{a_i=0}=g_{\mathrm{BK}}\big|_{\myGauss=-1}\,.
 \end{equation}
 In \eqref{21X24/1} the tensor field $\ringh_{AB}$ is an $(n-1)$-dimensional Einstein metric, and $\myGauss$ is the sign of  its  scalar curvature. 
 This also shows by continuity that the scalar curvature of the sections of the Killing horizons will be negative for small rotation parameters. 
 However, as in the previously known cases~\cite{Klemm:1997ea,Klemm:1998kd} it will not be constant.

 \subsection{Singularities}\label{sec: singularities}

 If $\bar m =0$ the metric is locally Anti-de Sitter in unusual coordinates~\cite{Gibbons:2004uw}, and we will demonstrate this explicitly in Section \ref{17X24f.1}.
 
 If $\bar m \ne 0$ we note that the length $\bar{g}_{\bar t\, \bar t}$ of the Killing vector $\partial_{\bar t}$, namely
 \begin{equation}\label{4XI24.1}
 \bar{g}_{\bar t\, \bar t} = \bW (1-\ellsquare \br^2) + \frac{2\bar{m}\br\bW^2}{\barUh}
 \,,
 \end{equation}
 tends to infinity as  zeros of $\barUh$ are approached.  
 Now, from \eqref{22X24.1} we have
 \begin{align}
 \label{22X.2}
 \bUh^{-1}  =   \bigg(1 + \sum_{i=1}^N \frac{ {\bmuisq}\baa_i^2}{\br^2+\baa_i^2}\bigg)^{-1} \prod_{i=1}^N (\bar{r}^2+\baa_i^2)^{-1} \,.
 \end{align}
 Therefore:
 
 \begin{enumerate}
 	\item Suppose that $\bar m\ne0 $ and at least one of the rotation parameters vanishes. 
 	Then the norm of the Killing vector $\partial_{\bar t}$ diverges, when    $\br\to 0$, as $\br^{-1}$. Hence the metric is $C^2$-inextendible across $\{\br = 0\}$.  
 	
 	When  all the parameters $\baa_i $ vanish we have $\bW = 1$ and $\bUh = \br^{2N }=\br^{n-1}$, in which case \eqref{4XI24.1} reduces to the Birmingham-Kottler formula (compare \eqref{1VIII24.2} above)
 	$$
 	\bar{g}_{\bar t\, \bar t} = 1-\ellsquare \br^2 + \frac{2\bm}{\br^{2n-2}}\,.
 	$$
 	\item The function  $\bUh^{-1}$ is bounded for all $\br $ if all the
 	$\baa_i$'s are non-vanishing, in which case $\bar{g}_{\bar t\, \bar t}$ remains finite  throughout the domain of definition of the metric.
 \end{enumerate}

 The remaining potential singularities of the tensor field $\bar g$ arise from  
 zeros of the determinant,  of the  $\red{\bmu^i}$'s,  and   of $\barVh-2\bm\br$.
 It turns out that all these can be got rid of  by introducing new coordinates.
 Indeed, the zeros of the determinant are determined in Appendix~\ref{s5XI24.1}, with the results there summarised in Proposition~\ref{P6XI24.1} below.
 The zeros of $\red{\bmu^i}$'s have already been shown to be innocuous polar-coordinates singularities at axes of rotation (cf.\ below \eqref{20XI24.1}), and do not require further attention.  
 
 We show in Section~\ref{ss9XI24.1} that $\bar g$ is analytically extendible through the zeros of  $\barVh-2 \bm\br$, which become Killing horizons in the extended metric.

 \subsection{The Kerr--Schild extension}\label{17X24f.1}
 
 Similarly to the Myers--Perry metrics, the metric \eqref{3VI24.f7} can be brought into Kerr--Schild form using a similar construction  as in \cite{Gibbons:2004uw}.
 
 To begin note that,  for $\br^2 \ne \ell^2$,  the Anti-de Sitter metric  
 can be written as  
 \begin{align}\label{1X24f.1}
 \barg _{AdS}=&\bar W(1 -\ellsquare \bar r^2)d\tau^2 +F d\bar r^2
 \nonumber
 \\
 &+\sum_{i=1}^N\frac{\bar{r}^2+\baa_i^2}{\bar{\Xi}_i}\bigl(\bmuisq  (d\red{\varphi^i})^2+\dbmuisq )
 - \bar{r}^2\red{(d\red{\bmu^{N+1}})^2}
 \nonumber
 \\
 &-\frac{\ellsquare }{\bar{W}(1 -\ellsquare \bar{r}^2)}\Bigl(\sum_{i=1}^N \frac{\bar{r}^2+\baa_i^2}{\bar{\Xi}_i}\red{\bmu^i} d\red{\bmu^i}
 - \br^2 \red{\bmu^{N+1}}d\red{\bmu^{N+1}}\Bigr)^2\,,
 \end{align}
 where $\bar W$ is as in \eqref{24VII24.f1} and
 \begin{equation}
 F=\frac{\bar r^2}{1 -\ellsquare \bar r^2}\left(
 \sum_{i=1}^{N}\frac{\bmuisq }{\bar r^2+\bar a_i^2} - \frac{(\red{\bmu^{N+1}})^2}{\bar r^2 }\right)
 =- \frac{1}{1 -\ellsquare \bar r^2}\left(
 1+
 \sum_{i=1}^{N}\frac{\baa_i ^2 \bmuisq }{ \bar r^2+\bar a_i^2 } \right)
 \,.
 \label{14XII24.1}
 \end{equation}
 To see that the metric \eqref{1X24f.1}-\eqref{14XII24.1} is locally AdS, one can use the coordinate transformation, defined implicitly by
 \begin{align}\label{26XI24f.1} 
 y^2\red{(\hat{\mu}^i)^2}=\frac{\bar r^2+\bar a_i^2}{\Xi_i}\bmuisq
 \,,
 \end{align}
 together with
 \begin{equation}\label{26XI24f.2}
 \red{(\hat\mu^{N+1})^2} -\sum_{i=1}^{N}\red{(\hat{\mu}^i)^2} =1\,.
 \end{equation}
 Notice that \eqref{26XI24f.1} and \eqref{26XI24f.2} mean
 \begin{equation}
 y^2= \bar r^2 \red{(\bmu^{N+1})^2} - \sum_{i=1}^{N}\frac{(\bar r^2+\bar a_i^2)\bmuisq}{\bar \Xi_i}
 \,,
 \end{equation}
 which implies
 \begin{equation}
 1 -\ellsquare y^2=\bar W(1 -\ellsquare \bar r^2) 
 \,.
 \label{24XI24.5}
 \end{equation}
 This  brings the metric \eqref{1X24f.1} into the form 
 \begin{align}
 \barg _{AdS}=&(1 -\ellsquare  y^2)d\tau^2 - \frac{dy^2}{1 -\ellsquare  y^2}
 +y^2
 \Big(
 \underbrace{
 	\sum_{i=1}^{N}
 	\big(
 	(d\red{\hat{\mu}^i})^2+\red{(\hat{\mu}^i)^2}(d\red{\varphi^i})^2
 	\big)
 	- (d\hat{\mu}^{N+1})^2
 }
 \Big)
 \,,
 \end{align}
 which is one of the  standard forms  for the AdS metric with negative cross-sectional curvature, keeping in mind that the underbraced terms need  to be restricted to the constraint hyperboloid \eqref{26XI24f.2}.

 One can use the $(\tau, \bar r, \red{\varphi^i}, \red{\hat\mu^i})$-coordinates to express the metric \eqref{3VI24.f78} as
 \begin{equation}\label{25XI24f.1}
 \barg =\barg _{AdS}
 +\frac{2\bar m\br}{\barUh}(k_\mu dx^\mu)^2\,,
 \end{equation}
 with $\barUh$ and $ \bar W$   as in \eqref{24VII24.f1}, where   $k_\mu$ is a null co-vector (with respect to both \eqref{1X24f.1} and \eqref{25XI24f.1})  given by
 \begin{align}\label{14X24 f.1}
 k_\mu dx^\mu&=\bar W d\tau +F d\bar r 
 +\left(
 \sum_{i=1}^{N}\frac{ \bar a_i(\hat{\mu}^i)^2}{\bar \Xi_i}d\red{\varphi^i}
 \right)\,,
 \\
 k^\mu\partial_\mu&=\frac{1}{1 -\ellsquare \bar r^2}\frac{\partial}{\partial\tau}+\frac{\partial}{\partial\bar r} +\sum_{i=1}^{N}\frac{\bar a_i}{\bar r^2+\bar a_i^2} \frac{\partial}{\partial \red{\varphi^i}} 
 \,.
 \end{align}

 The metric in the original form \eqref{3VI24.f7} is recovered from \eqref{25XI24f.1} using the coordinate transformation 
 \begin{equation}
 d\tau=d\bar t- 
 \frac{2\bar m\br\, d\bar r}{(1 -\ellsquare \bar r^2)(\barVh -2 \bar m\br)}
 \,,\quad
 d\red{\varphi^i}=d\red{\phi^i} %- \sigma^{-2}\malphasquare \bar a_i d \bar t 
 -
 \frac{2\bar m \br \bar a_i\, d\bar r}{(\bar r^2+\bar a_i^2)(\barVh -2 \bar m\br)}
 \,.
 \label{12XI24.1}
 \end{equation}
 Not unexpectedly, the coordinate transformation \eqref{12XI24.1} is singular at zeros of both $1 -\ellsquare \bar r^2$  and $\barVh -2 \bar m\br$: the Kerr-Schild coordinates provide an extension of the original form of $\bar g$ across the zeros of $\barVh -2 \bar m\br$, while the original form of the metric provides an extension of the Kerr-Schild form of the metric across the zeros of 
 $ 1 -\ellsquare \bar r^2$.

 \subsection{Asymptotics}
 \label{ss28VII24.4}

 The behaviour of the metric at large distances can be directly read from \eqref{3VI24.f78}. 
 For large $\br$ we find
 \begin{align}\label{11XI24.21}
 \bar g  =&
 -\bar{W} \ellsquare\bar{r}^2
 \big(1+O(\br^{-2})
 \big)
 d\bar{t}^2
 +\frac{\ell^2
 	\big(1+O(\br^{-2})
 	\big) 
 	d\bar{r}^2}{ \br^2}
 +O(\br^{1-2N} )(\bar{W}d\bar{t} 
 +  \sum_{i=1}^N \frac{\baa_i \bmuisq d\red{\phi^i}}{\bar{\Xi}_i}   
 \biggr)^2
 \nonumber
 \\
 & 
 +\sum_{i=1}^N\frac{\bar{r}^2+\baa_i^2}{\bar{\Xi}_i}\bigl(\bmuisq\red{(d\phi^i)^2}+\dbmuisq ) 
 -
 \bar{r}^2 
 \frac {\Big( \tfrac{1}{2}\sum_{i=1}^{N} d(\bmuisq )\Big)^2}{1 +   \sum_{i=1}^{N} \bmuisq}  
 - \frac{ \ell^2- \bar{r}^2}{4 \bar W}
 ( d\bar W)^2\,.
 \end{align}
 Introducing a new coordinate
 $$
 x: = 1/\bar r
 \,,
 $$
 we see that the tensor field  $x^2 \bar g$ extends analytically  across  a conformal boundary at infinity
 $$
 \Scri:= \{x = 0\}
 \,,
 $$
 and is non-degenerate in a neighborhood of $\Scri$. 
 As such, the metric $x^2 \bar g$ induces on $\Scri$ the tensor field 
 \begin{align}\label{11XI24.22}
 &
 -\bar{W} \ellsquare 
 d\bar{t}^2 
 +\sum_{i=1}^N\frac{\bmuisq\red{(d\phi^i)^2}}{\bar{\Xi}_i}
 \underbrace{
 	+\sum_{i=1}^N\frac{\dbmuisq }{\bar{\Xi}_i}
 	+ \frac{ ( d\bar W)^2}{4 \bar W}
 	-  
 	\frac {\Big( \tfrac{1}{2}\sum_{i=1}^{N} d(\bmuisq )\Big)^2}{1 +   \sum_{i=1}^{N} \bmuisq}   
 }
 \,.
 \end{align}
 It follows that  $\bar t$ is a time function for all $\br$ large enough, but 
 the signature of the tensor field \eqref{11XI24.22} is not  apparent.  
 In particular, the signature of the underbraced $\red{\bmu^i}$ sector, which we denote $\bar{g}_{ij}$, is not obviously Riemannian.
 But taking the limit $\br \to \infty$ in \eqref{25VII24f.7} below gives
 \begin{align}\label{25VII24f.7hr}
 \lim_{\br\to\infty} \bar{r}^{-2N} \det(\bar g_{ij})  
 = &\lim_{\br\to\infty} 
 \frac{1}{\bar{W}\,\red{(\bmu^{N+1})^2}}
 \left(
 1  + \sum_{i=1}^{N}\frac{\baa_i^2 \bmuisq}{\bar{r}^2+\baa_i^2}
 \right )
 \prod_{i=1}^{N} \frac{1+\baa_i^2\bar{r}^{-2} }{\bar{\Xi}_i}
 \nonumber
 \\    
 = &
 \left(\big(1 +  \sum_{i=1}^{N }
 \frac{ \bar a_i^2\bmuisq}{\ell^2 \bar\Xi_{i}}\big)\big(	1 +   \sum_{i=1}^{N} \bmuisq\big) \prod_{i=1}^{N} \bar{\Xi}_i
 \right)^{-1}
 \,,
 \end{align} 
 which has no zeros.
 Since the tensor field \eqref{11XI24.22} obviously has Lorentzian signature at $\{\red{\mu^i}=0\}$ when using the coordinates \eqref{20XI24.1}, we conclude that \eqref{11XI24.22} has Lorentzian signature everywhere.
 
 While it is not immediately apparent, the metric \eqref{11XI24.22} is conformal to a hyperbolic version of the Einstein cylinder. 
 This follows from  the Kerr-Schild form of the metric, in which the metric $y^{-2} \bg$ extends smoothly across $\{z=0\}$, where $z:= 1/y$, with the extension inducing on $\{z=0\}$ the metric 
 \begin{align}
 -\ellsquare  d\tau^2 +   g_{\mcH}
 \,,
 \end{align}
 where $g_{\mcH}$ is the metric on hyperbolic space.

 \subsection{Extensions, Killing horizons}
 \label{ss9XI24.1} 
 
 We will say that the metric is \emph{fully rotating} if
 \begin{equation}\label{1XII24.1}
 \bm\,  \baa_1\cdots \baa_N \ne 0
 \,.
 \end{equation}  
 We wish to show that, in the  fully-rotating case, the tensor field \eqref{3VI24.f78} can be extended through all the remaining potentially singular sets, namely zeros of $1-\ellsquare  \bar r^2$ in the Kerr--Schild form of the metric, and  zeros of $  \barVh -2\bar m\br $ in \eqref{3VI24.f78}.
 
 \subsubsection{$1-\ellsquare  \bar r^2$} 
 \label{sss11XI24.2}
 In the  fully-rotating case the tensor field \eqref{3VI24.f78} is manifestly regular near those zeros of $1-\ellsquare  \bar r^2$ which are distinct from zeros of  $  \barVh -2\bar m\br $.  
 We will show shortly that these zeros are distinct under \eqref{1XII24.1}.  
 
 Possible concerns about the signature of \eqref{3VI24.f78} at zeros of $1-\ellsquare  \bar r^2$ are dispelled by Proposition~\ref{P6XI24.1}.  
 So, as already mentioned, the original form of the metric provides an extension of the Kerr--Schild form of the metric across $\{\br = \pm \ell\}$.

 \subsubsection{ $  \barVh -2\bar m\br $} 
 \label{sss9XI24.1}
 
 Recall that
 \begin{align}\label{28VII24.11} 
 \barVh 
 &
 =
 -(1-\ellsquare \bar{r}^2 )
 \prod_{i=1}^\red{N}(\bar{r}^2+\baa_i^2
 )\,,
 \end{align}
 thus $  \barVh -2\bar m\br $ vanishes when $\br$ is a root of the equation  
 \begin{align}\label{28VII24.11a} 
 (1-\ellsquare \bar{r}^2)
 \prod_{i=1}^\red{N}(\bar{r}^2+\baa_i^2
 )
 = 
 -
 2\bar m  {\br}
 \,.
 \end{align}
 When $\baa_i\neq 0$, the case $\bar m \bar r =0$ in \eqref{28VII24.11a}  is only possible if $\bar m=0$. 
 In this  case the rotation parameters are actually irrelevant in that they can be got rid of by a coordinate transformation: 
 Indeed, it follows from Section~\ref{17X24f.1} that the metric $\bar g$ is 
 then locally isometric to the Anti-de Sitter metric. 
 Changing $\br $ to its negative if necessary, it is standard (cf., e.g., \cite[Chapter~6]{ChBlackHoles}) to construct an extension where 
 $\br >  0$,  and in which the hypersurface  $\{\bar r =  \ell\}$ is a bifurcate Killing horizon. 
 
 In the fully-rotating case \eqref{1XII24.1}, the roots of $\barVh - 2\bm\br$  occur away from the set  $\{\br = \pm \ell\}$. 
 It then follows that the Kerr--Schild form of the metric \eqref{25XI24f.1}-\eqrefl{14X24 f.1} provides an analytic extension across each of the roots of $\barVh - 2\bm\br$. 
 Further extensions are constructed in Section~\ref{s15XI24.1}. 

 We claim that,again in the fully-rotating case, the function $\barVh - 2\bm\br$   has at least two zeros $\br_\pm$  satisfying  
 \begin{equation}\label{9XI24.14}
 \left\{
 \begin{array}{ll}
 \br_-<-\ell<0<\br_+<\ell
 \,, & \hbox{$\bar m <0$;}
 \\
 -\ell<  \br_- <0<\ell< \br_+ 
 \,, & \hbox{$\bar m >0$.}
 \end{array}
 \right.
 \end{equation}
 To see this, note that the zeros of $\barVh - 2 \br \bar m$ are  roots of the polynomial
 \begin{align}\label{9XI24.11} 
 F(\br):= (\ell^2 -\br^2)
 \prod_{i=1}^\red{N}(\br^2+\baa_i^2)
 +
 2\ell^2\bar m \br
 \,.
 \end{align}
 Assuming \eqref{1XII24.1} we have 
 \begin{equation}\label{9XI24.12}
 F(0) >0\,,
 \quad
 F(\pm \ell) = \pm 2 \ell^3 \bar m 
 \,,
 \quad
 \lim_{\br\to\pm\infty} F = - \infty\,.
 \end{equation}
 Consider the possibility that $\bar m>0$. 
 Then $F(1)>0$, and since $F\to_{\bar r\to\infty} -\infty$ there must be a zero, say  $\bar r_+$, of $F$ in $(1,\infty)$. 
 Similar arguments apply to the remaining cases in \eqref{9XI24.14}.
 
 When $\bar m$ vanishes but no  rotation parameter  does, the function $F$ has exactly two zeros, at $r = \pm \ell$. 
 Continuity shows that it will have exactly two zeros in the fully-rotating case when $\bm$ is small enough. 
 In Figure~\ref{F11XI24.1} we give examples of parameters where $F$ has three or four zeros.  
 In Appendix~\ref{app9XI24.1} we show that $F$ can have at most four zeros.
 \begin{figure}
 	\centering
 	\includegraphics[scale=.6]{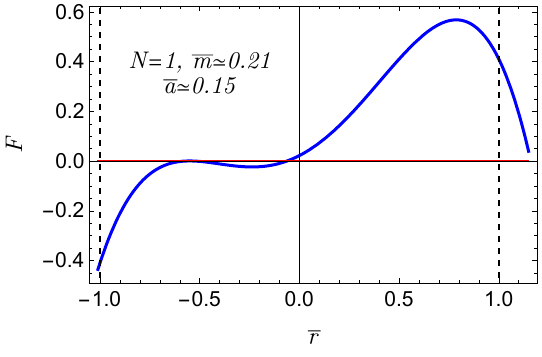}
 	\includegraphics[scale=.6]{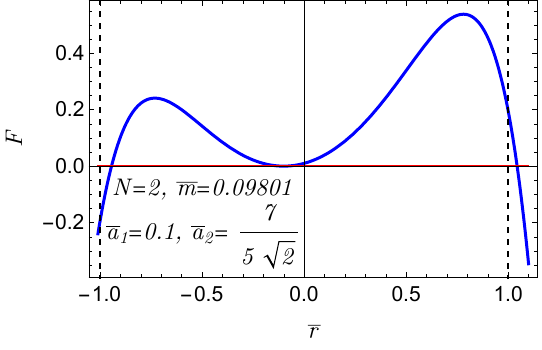}
 	\includegraphics[scale=.6]{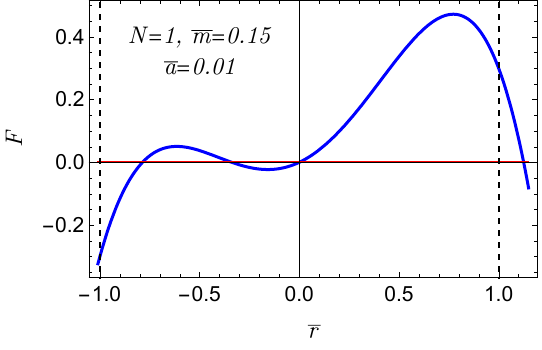}
 	\includegraphics[scale=.6]{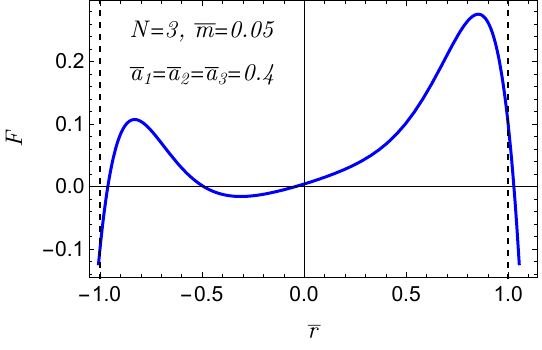}
 	\caption{Examples of  the function $F$ with three (top) or four (bottom) zeros. In all plots, $\ell = 1$.}\label{F11XI24.1}
 \end{figure}

 We have
 $$
 g^{\bar r \bar r} =  \frac {\barVh - 2\bar m \br   }{\barUh } 
 \,,
 $$
 which shows that each smooth connected subset of 
 $$
 \Hor:=\{ \barVh= 2\bm\br   
 \}   
 $$
 is null, with  the Killing vectors $\partial_t$ and $\partial_{\phi^i}$ tangent to $\Hor$.  
 If $\br_H$ is a zero of $\barVh - 2 \bar m\br$ then the following linear combination of Killing vectors
 \begin{equation}\label{28VII24.5}
 K:= \partial_\tau + \sum_i \Omega_i\partial_{\red{\varphi^i}}\,,
 \quad
 \mbox{where}
 \
 \Omega_i:=\frac{\baa_i(1-\ellsquare \bar r_H^2)}{\bar r_H^2+\bar a_i^2}
 \end{equation}
 is null on $\Hor$.  
 We conclude that, in the fully-rotating case, each connected component of
 $$
 \Hor\cap \{K\ne 0\}
 $$ 
 is a Killing horizon.

 \subsection{Surface gravity}
 \label{ss15XI24.1}
 
 Since the vector field $K$ is null on the horizon  we have $K^\mu\nabla_\mu K_\nu=\kappa K_\nu$ there, where $\kappa$ can also be determined from $\kappa^2=\frac12\nabla_\mu K_\nu \nabla^\mu K^\nu$. 
 As is well known, $\kappa$ is constant on each connected component of $\Hor$ in vacuum, and is called \emph{surface gravity}. 
 The calculation of $\kappa$ is conveniently done in Kerr--Schild coordinates, as $\bar g$ is not defined on the horizons in Boyer--Lindquist type coordinates.  
 A lengthy calculation gives 
 \begin{equation}
 \pm\kappa=\br_H(\ell^{-2}\br_H^2-1)\left(\sum_{i=1}^N \frac{1}{\br_H^2+\baa_i^2}+\frac{1}{2\br_H^2}\right)+\frac{1}{\br_H}
 \,;
 \label{15XI24.1}
 \end{equation}
 a resolution of the sign-ambiguity requires a choice of time orientation. 
 Note that degenerate horizons, i.e.\ horizons with $\kappa=0$, can only occur in the region where $|\br| < \ell$, hence for $\bm>0$ the outermost horizon is always non-degenerate.

 \subsection{Signature}
 \label{ss28VII24.1}
 
 It follows from Section~\ref{ss28VII24.4} that $\bar g$ has Lorentzian signature at large distances.
 
 Keeping in mind that at the axes of rotation, i.e.\ $\red{\mu^i}=0$, the tensor field $\bar g$  is regular in suitable coordinates, the calculations of  Appendix~\ref{s5XI24.1} show that:
 
 \begin{Proposition}
 	\label{P6XI24.1}
 	The tensor field  $\bar g$ of \eqref{3VI24.f78} is non-degenerate on every connected component of the set 
 	\begin{equation}\label{6XI24.5}
 	\{\barVh\ne 2\bar{m} \br\}
 	\,.
 	\end{equation}
 	\qed
 \end{Proposition}

 It follows immediately that $\bar g$ has Lorentzian signature on those connected components of the set $\{\barVh\ne 2\bar{m} \br\}$ on which $\bar r\to \pm \infty$.
 
 Now, the analysis of Section~\ref{sss9XI24.1} shows that the signature does not change across the zeros of $\barVh-2\bar{m} \br $, when $\bar g$ is extended across these zeros using the Kerr--Schild form of the metric.  
 Further, it follows from \eqref{24XI24.5} that $\br$ is a smooth function on the Kerr--Schild extension. 
 Its level sets away from the zeros of $\barVh-2\bar{m} \br $ are all diffeomorphic to $ \R^{2N+1}$ on the manifold underlying the original form \eqref{3VI24.f78} of the metric. 
 Furthermore, \eqref{24XI24.5} shows that 
 \begin{enumerate}
 	\item the sets $\{\br=\pm \ell\}$ and $\{y=\pm \ell\}$ coincide, with 
 	\item
 	$
 	0\ne dy|_{y=\pm\ell}  =\bar Wd\br
 	$.
 \end{enumerate}
 Hence the level sets of $\br$ are also diffeomorphic to $ \R^{2N+1}$ near the zeros  of $\barVh-2\bar{m} \br $  in the Kerr--Schild extension.  
 
 We conclude that the extension of  the metric \eqref{3VI24.f78}
 through each connected component of the set $\{\barVh-2\bar{m} \br=0 \} $, as  obtained by using the Kerr--Schild coordinates, provides a smooth Lorentzian metric defined on  
 \begin{equation}\label{14XII24.2}
 \mcM_{\mathrm{KS}}:=\{ \br\in \R\}\times \R^{2N+1}
 \,.
 \end{equation}
 The function $\br$ is a globally defined smooth function on $ \mcM_{\mathrm{KS}}$, 
 while the coordinates $(\bt, \red{x^i},\red{y^i} )\in \R^{2N+1}$, where $(\red{x^i},\red{y^i} )$ are as in \eqref{20XI24.1},  are smooth coordinates away from  the zeros of $\barVh-2\bar{m} \br $, and have to be patched together using  Kerr--Schild coordinates across the horizons.

 \subsection{Causality}
 \label{ss24XI24.1}

 \subsubsection{Stable causality}

 Recall that a spacetime $(\mcM,g)$ is \emph{stably causal} if there exists on $\mcM$ a function with timelike gradient; and a function is a \emph{time function} if its gradient is {timelike past pointing}. 
 
 We have
 $$
 g(\nabla r, \nabla r)=g^{\bar r \bar r} =  \frac {\barVh - 2\bar m \br   }{\barUh } 
 \,,
 $$
 which shows that each connected component of the set $\{\barVh - 2\bar m \br< 0\}$ is stably causal,  with either $\br$ or $-\br$ being time functions there.  
 
 Since the metric reduces to a BK metric when the rotation parameters vanish, the obvious candidate for a time function  on the set
 $\{\barVh - 2\bm \br>0\}$   is  the coordinate $\bt$, or its negative.
 Now,  it follows from the  block diagonal nature of the metric that 
 to find $g^{\bar t\, \bar t}$ 	we need to invert the 
 $x^\alpha=(\bar t,\red{\phi^i})$-sector of the metric. 
 This is most conveniently done using  the ADM decomposition of this sector of the metric:
 \begin{align}
 \hat{g} :=\ &
 \bar{W}(1-\ellsquare\bar{r}^2)d\bar{t}^2 
 +\frac{2\bar{m}\br}{\barUh}
 \biggl(\bar{W}d\bar{t} 
 +  \sum_{i=1}^N \frac{\baa_i \bmuisq d\red{\phi^i}}{\bar{\Xi}_i}   
 \biggr)^2 
 +\sum_{i=1}^N
 \frac{\bar{r}^2+\baa_i^2}{\bar{\Xi}_i} 
 \bmuisq\red{(d\phi^i)^2} 
 \nonumber
 \\       
 =: \
 & 
 \pm N^2 d\bar t^2 + h_{ij}(d\red{\phi^i} + N^i d\bar t) (d\red{\phi^j}+ N^j d\bar t)
 \,.\label{15XI24.4}
 \end{align}
 We have, from the Equations \eqref{20XI24.f2}-\eqref{20XI24.f3}, \eqref{5XII24.f1}, and \eqref{20XI24.f5} of Appendix~\ref{s5XI24.1}, 
 \begin{align}\label{16XI24.3}
 h:= h_{ij} d\red{\phi^i} d\red{\phi^j} 
 = \ & 
 \Big(
 \frac{\bar{r}^2+\baa_i^2}{\bar{\Xi}_i} 
 \bmuisq \delta_{ij} 
 +  \frac{2\bar m  \bar r  \baa_i\baa_j  \bmuisq  \red{(\bmu^j)^2}}
 {\bar U \bXi_i \bXi_j }
 \Big) d\red{\phi^i} d\red{\phi^j} 
 \,, 
 \\
 N^j = \ & %h^{ij} 
 %\frac{2\bar m \bar r \bar W\baa_i \bmuisq }{\bar U \bXi_i}
 {2\bm\br} \bW
 \left({\barUh}+{2\bm\br}\,
 \sum_{i=1}^N\frac{\baa_i^2\bmuisq}{(\br^2+\baa_i^2)\bXi_i} \right)^{-1} \frac{\baa_j}{\br^2+\baa_j^2}
 \,, \label{20XI24.f7}
 \\
 \pm N^2 = \ & -(\barVh-2\bm\br)\frac{\bW}{\left(1+\frac{2\bm\br}{\barUh}\sum_{i=1}^N\frac{\baa_i^2\bmuisq}{(\br^2+\baa_i^2)\bXi_i}\right)\prod_{i=1}^{N}(\br^2+\baa_i^2)}
 %\bar{W}(1-\ellsquare\bar{r}^2) 
 %   + \frac{2\bar{m}\br}{\barUh} \bar{W}^2
 %   -
 %   % \sum_{i,j =1}^N 
 %   h_{ij}N^i N^j
 \,,
 \label{16XI24.91}
 \end{align}
 Hence, away from zeros of $(\barVh-2\bm \br)$, 
 \begin{align}\label{28VII24.41}
 \bg^{\bar t\, \bar t}   
 &
 = \pm N^{-2}=
 - \frac{\prod_{i=1}^N(\bar{r}^2+\baa_i^2)}{\bar W (\barVh-2\bm \br)}
 \Bigg( 1 
 + 
 \frac{2\bm  \br }{\barUh } \sum_{i=1}^{N}\frac{\baa_i ^2\bmuisq}{\Xi_i (\bar{r}^2+\baa_i^2)} 
 \Bigg)
 \\
 &=
 - \frac{1}{\bar W (\barVh-2\bm \br)}
 \Bigg(   \prod_{i=1}^N(\bar{r}^2+\baa_i^2)
 + 
 \frac{2\bm  \br }
 {1+ \sum_{i=1}^{N}\frac{\baa_i ^2\bmuisq}{\bar{r}^2+\baa_i^2}}
 \sum_{i=1}^{N}\frac{\baa_i ^2\bmuisq}
 {\Xi_i
 	(\bar{r}^2+\baa_i^2)  }
 \Bigg)
 \label{25XI24.23}
 \\ 
 &= - \frac{1}{\bar W (\barVh-2\bm \br)\big(1+ \sum_{i=1}^{N}\frac{\baa_i ^2\bmuisq}{\bar{r}^2+\baa_i^2}\big)}
 \times 
 \nonumber\\
 &
 \quad
 %\times
 \left( 
 \prod_{i=1}^N(\bar{r}^2+\baa_i^2)\big(1+ \sum_{i=1}^{N}\frac{\baa_i ^2\bmuisq}{\bar{r}^2+\baa_i^2}\big)
 -\frac{2\bm\br\ell^2}{\br^2-\ell^2} 
 \left[ \sum_{i=1}^{N}
 \frac{\baa_i ^2\bmuisq} {\bar{r}^2+\baa_i^2 }
 -\frac{\baa_i ^2\bmuisq} {\ell^2\Xi_i}
 \right]
 \right)
 \nonumber
 % \label{25XI24.23}
 \\ 
 &= - \frac{\ell^2}{(\br^2-\ell^2)\bar W (\barVh-2\bm \br)\big(1+ \sum_{i=1}^{N}\frac{\baa_i ^2\bmuisq}{\bar{r}^2+\baa_i^2}\big)}
 \left( 
 (\barVh-2\bm\br)\left(1+ \sum_{i=1}^{N}\frac{\baa_i ^2\bmuisq}{\bar{r}^2+\baa_i^2}\right)
 +2\bm\br\bW
 \right)
 \label{13XII24f.1}
 \,.
 \end{align}
 Strictly speaking, the last two equations are only valid if $\br^2 - \ell^2\ne 0$. However, it is immediate from \eqref{25XI24.23} that \eqref{13XII24f.1} has a finite limit when $|\br|\to\ell$ in the fully rotating case.
 
 Since $\bW>1$ and $\bar U>0$, Equation~\eqref{28VII24.41} shows that
 $\bar\nabla \bt$ is manifestly timelike  on the set     
 $$\{\barVh - 2 \bm \br > 0\}\cap \{ \bm \br >0\}
 \,.
 $$ 
 It thus  remains to determine the sign of  
 \begin{equation}\label{13XI24.1}
 (*):=  \prod_{i=1}^N(\bar{r}^2+\baa_i^2)
 +2\bm  \br
 \underbrace{
 	\frac{   \sum_{i=1}^{N}\frac{\baa_i ^2\bmuisq}
 		{\Xi_i
 			(\bar{r}^2+\baa_i^2)  } }
 	{1+ \sum_{i=1}^{N}\frac{\baa_i ^2\bmuisq}{\bar{r}^2+\baa_i^2}}
 }_{\in (0,1)} 
 \,.
 \end{equation}
 assuming $\bm \br <0$ and 
 \begin{align}\label{13XI24.2}
 0< \barVh - 2 \bm \br
 = -(1
 & 
 -\ellsquare \bar{r}^2)
 \prod_{i=1}^N(\bar{r}^2+\baa_i^2) - 2 \bm \br 
 \nonumber
 \\ 
 &
 \phantom{xxxx}
 \Longleftrightarrow
 \quad
 (1-\ellsquare \bar{r}^2)\prod_{i=1}^N(\bar{r}^2+\baa_i^2) < 2 |\bm \br|
 \,.
 \end{align}
 We have therefore, for a fully rotating configurations:
 
 \begin{enumerate}
 	\item 
 	At any point with $\br \ne 0$ we can make \eqref{13XI24.1} negative  while maintaining \eqref{13XI24.2}  by making $\bar m$ negative large enough. 
 	Thus there exist sets of parameters and subsets of $\{\barVh - 2 \bm \br > 0\}$ where $\bar \nabla \bar t$ is \emph{not} timelike.  
 	We will show in Section~\ref{sss11XI24.1} that stable causality is violated whenever $(*)$ fails to be positive.
 	
 	\item  On the intersection of the axis $\{\red{\bmu^i}=0\}$ with the region $\{\barVh - 2 \bm \br > 0\}$, the function $\bt$ is a time function, after the time-orientation has been suitably chosen. 
 	
 	\item 
 	For any set of mass and rotation parameters, the function $\bar t$ is a time function for $|\br|$ large enough.
 	\item Similarly $\bar t$ is a always a time function for $|\br|$ small enough. 
 	\item For every set of rotation parameters there exists a range of masses for which $\bar t$ is a time function throughout every connected component of $\{\barVh - 2 \bm \br > 0\}$. Indeed, let
 	$$
 	a:=   \prod_{i=1}^N \baa_i^2
 	\,.
 	$$
 	Then  for $ |\br|\le \rho  $ we have
 	$$
 	(*)    \ge a  - 2 |\bm| \br   \ge a  - 2 |\bm| \rho  
 	\,,
 	$$
 	which will be positive if $|\bm|< a/(2\rho)$. On the other hand, for $|\br|\ge \rho$ we have 
 	$$
 	(*)    \ge \br^{2N}  - 2 |\bm \br| = |\br|( |\br|^{2N-1}  - 2 |\bm | )
 	\ge \rho  ( \rho^{2N-1}- 2 |\bm|)  
 	\,,
 	$$
 	which will be positive if $|\bm| < \rho^{2N-1}/2$. Choosing $\rho$ so that $  a/(2\rho)=\rho^{2N-1}/2 $, namely $\rho = {a}^{1/(2N)}$,  we find
 	\begin{equation}\label{13XI24.1nop}
 	(*)>0  \ \mbox{for}\  |\bm| < \frac{ a^{\frac{2N-1}{2N}}}{2}
 	\,.
 	\end{equation}
 \end{enumerate}
 
 We have therefore proved:
 
 \begin{Proposition}
 	\label{P15xi24.2}
 	Suppose that 
 	$$
 	2 |\bm| <  \prod_{i=1}^N |\baa_i|^{\frac{2N-1}{N}} 
 	\,.
 	$$
 	Then each connected component of the set $\{\bar V \ne 2 \bar m \bar r\}$ is stably causal. 
 	\qed
 \end{Proposition}

 \subsubsection{Closed or almost closed causal curves, time machine}
 \label{sss11XI24.1}
 A minimum requirement for a well behaved spacetime is non-existence of closed causal curves. 
 Obvious candidates for such curves are the orbits of the Killing vectors $\partial_{\red{\phi^i}}$. 
 In the region $\bar g^{\bt\,\bt}<0$ the level sets of $\bt$ are spacelike, and so are therefore the integral curves of the $\partial_{\red{\phi^i}}$'s. 
 However, since
 \begin{equation}\label{1VIII24.11}
 \bar\fourg(\partial_{\red{\phi^i}},\partial_{\red{\phi^i}}) =
 \frac{2 \bar m\br}{\barUh}
 \left(\frac{\bar a_i \bmuisq }{ \bar \Xi_i}\right)^2+\frac{(\bar r^2+\bar a_i^2)\bmuisq}{\bar \Xi_i} 
 \end{equation}
 we see that at any point with $\br \ne 0$ and for any collection of rotation parameters we can choose $\bar m$ with $\bar m \bar r <0$   so that one, 
 or more (possibly all), of the $\partial_{\red{\phi^i}}$'s is timelike or null. 
 The orbits of such $\partial_{\red{\phi^i}}$'s are closed causal curves.
 
 More generally, consider the matrix of scalar products of the Killing vectors  $\partial_{\red{\phi^i}}$, 
 \begin{equation}\label{14XII24.3}
 h \equiv
 \big(h_{ij}:= \bg(\partial_{\red{\phi^i}},\partial_{\red{\phi^j}})
 \big)
 \,.
 \end{equation}
 %.  
 As shown in Equation~\eqref{25XI24.1} of Appendix~\ref{s5XI24.1},  we have  
 \begin{equation}
 \det h= \Bigg(1+\frac{2\bm\br}{\barUh}
 \sum_{i=1}^N\frac{\baa_i^2\bmuisq}{(\br^2+\baa_i^2)\bXi_i}
 \Bigg)\prod_{i=1}^{N}\frac{(\br^2+\baa_i^2)\bmuisq}{\bXi_i}\,.
 \label{25XI24.21}
 \end{equation}
 Using 
 \eqref{30XI24.11} and \eqref{25VII24f.8ny} one finds
 \begin{equation}
 -\bW(\barVh-2\bm\br) \prod_{i=1}^{N}\frac{\bmuisq}{\bXi_i}
 = \det \hat{g}=\pm  N^2 \det h 
 = \frac{\det h}{\hat g(\hat \nabla \bt, \hat  \nabla  \bt)}
 = \frac{\det h}{\bar g(\bnabla \bt, \bnabla \bt)}
 \,.
 \label{30XI24.12}
 \end{equation}
 So $\bnabla \bt$ is timelike at  $p\in\{\barVh-2\bm\br > 0\}$ if and only if
 the matrix $h$ defined in \eqref{14XII24.3} is positive definite at $p$. Alternatively, one can compare \eqref{25XI24.21} with \eqref{16XI24.91} to reach the same conclusion.  
 
 Let $\mcN\approx \mathbb{T}^N$ denote any principal orbit of the action of the isometry group generated by  the Killing vectors fields $\partial_{ \red{\phi^i}}$, and consider the pullback, which we denote by $h_\mcN$, of  $h=h_{ij} d\red{\phi^i} d\red{\phi^j}$ to $\mcN$. When non-degenerate,  $h_\mcN$ is the metric induced by $\bg$ on $\mcN$. 
 Since $h_\mcN$ is Riemannian on the set $ \{\barVh-2\bm\br < 0\}$, which is stably causal as $\bnabla \br$ is timelike there, we have proved:
 
 \begin{Proposition}
 	\label{P30XI24.1}
 	Every  connected component of the set
 	$$\{\bar V \ne 2 \bm \br \}\cap \{\det h_\mcN >0\}
 	\,,
 	$$
 	where   $h_\mcN$ denotes the tensor field induced on the orbits of the group generated by the $\partial_{\red{\phi^i}}$'s, is stably causal. 
 \end{Proposition} 
 
 Suppose, next, that $h_\mcN $ is degenerate, then there exists a combination of the Killing vectors $\partial_{\red{\phi^i}}$, say $Y$, which is null. 
 Then the orbit of $Y$ is a null curve, which has a non-empty $\omega$-limit set by compactness of $\mcN$. Therefore the connected component of the set 
 $\{\bar V \ne 2 \bm \br \}$ 
 which contains $\mcN$ cannot be strongly causal, hence it cannot be stably causal either. 
 
 The last possibility is that $h_\mcN$ has Lorentzian signature. 
 Then $(\mcN,h_\mcN )$ is a compact Lorentzian manifold, and therefore contains a closed timelike curve by a theorem of Geroch (cf., e.g., Proposition~3.7.1 in \cite{ChBlackHoles}).
 Hence causality is violated at $\mcN$. By an argument of Carter (see \cite[Section~4.6.4]{ChBlackHoles}) there exists a causal curve through every point of $J^+(\mcN)\cap J^-(\mcN)$. 
 We expect that this last set coincides with the connected component of the set $\{\bar V \ne 2 \bm \br \}$ containing $\mcN$, but we have not attempted to establish this.

 From what has been said and from \eqref{28VII24.41} it follows that:  
 \begin{Proposition}
 	\label{P24XI24.1}
 	1.  All connected components of the set $\{\bar V- 2 \bm \br <0 \}$ are stably causal, with $\bnabla \br $ timelike there.

 	2.  A connected component of the set $\{\bar V- 2 \bm \br > 0 \}$ is stably causal if and only it holds there that  
 	\begin{align}  
 	1+\frac{2\bm\br}{\barUh}\sum_{i=1}^N\frac{\baa_i^2\bmuisq}{(\br^2+\baa_i^2)\bXi_i} 
 	>0
 	\,,
 	\label{16XI24.91b}
 	\end{align} 
 	and then $\bnabla \bt$ is  timelike there.
 	
 	3. The inequality \eqref{16XI24.91b} is satisfied  on 
 	the sets  $\{\bar V- 2 \bm \br > 0\}\cap \{ \bm \br >0\}$ and 
 	$\{\bar V- 2 \bm \br <0\}$.
 	\qed
 \end{Proposition}

 We define the \emph{time-machine set} as the set where closed causal curves lying entirely on an orbit  of the connected component of the isometry group of $\bg$ occur.
 We note that the pull-back, by the projection map considered in Section~\ref{s15XI24.1} below, of the Minkowskian time function, provides a continuous time function on  that part of the space-time from which the time-machine set has been removed. 
 In particular, for those values of parameters for which the matrix $\bg(\partial_{\phii}, \partial_{\phij})$ is positive definite everywhere, both the Kerr-Schild extension  and the remaining extensions considered in Section~\ref{s15XI24.1}  are stably causal.

 \subsection{Sections of the horizon}
 \label{ss28VII24.2}
 
 The question arises about the geometry of the cross-sections of the horizons.  For this consider the intersections, which we denote by $\mcN_{\tau, \br}$, of the level sets of the coordinates $\tau$ and $\br$ of the Kerr-Schild form \eqref{25XI24f.1} of the metric, with the induced metric which we   denote by $\ringh_\KS$: 
 \begin{align}
 \ringh_\KS
 =&  \sum_{i=1}^N\frac{\bar{r}^2+\baa_i^2}{\bar{\Xi}_i}
 \bigl(\bmuisq  (d\red{\varphi^i})^2+\dbmuisq )
 - \bar{r}^2 
 \frac {\Big( \tfrac{1}{2}\sum_{i=1}^{N} d(\bmuisq )\Big)^2}{1 +   \sum_{i=1}^{N} \bmuisq}  
 \nonumber
 \\
 &
 -\frac{\ellsquare  (1-\ellsquare \bar{r}^2)}{\bar{W}}
 \Bigl(
 \sum_{i=1}^N \frac{\baa_i^2}{\bar{\Xi}_i}\red{\bmu^i} d\red{\bmu^i} 
 \Bigr)^2 
 +\frac{2\bar m\br}{\barUh}
 \Big(
 \sum_{i=1}^{N}\frac{ \bar a_i\bmuisq}{\bar \Xi_i}d\red{\varphi^i}
 \Big)^2\,.
 \label{19XII24.2}
 \end{align}
 When $\br_H$ is a solution of the equation $ \bar V(\br_H) = 2 \bm \br_H$, then $\ringh_\KS$ is the metric induced on the cross-sections of the horizon.
 As shown in Section~\ref{ss28VII24.1}, the signature of the metric is Lorentzian, which implies that $\ringh_{KS}$ is Riemannian on all horizons. 
 
 \subsubsection{Negative Ricci curvature}
 
 Let us show that $\ringh_{KS}$ has a smooth conformal completion at infinity. For this we use  coordinates as in \eqref{16XI24.11a}, together with
 \begin{align}
 \label{22X24.xc1}
 &
 \bar W=1 +  \sum_{i=1}^{N }
 \frac{ \bar a_i^2\bmuisq}{\ell^2 \bar\Xi_{i}} 
 \,,
 \qquad 
 \bUh    =  \bigg(1
 + \sum_{i=1}^N \frac{ {\bmuisq}\baa_i^2}
 {\br^2+\baa_i^2}\bigg) \prod_{i=1}^N (\bar{r}^2+\baa_i^2) 
 \,,
 &
 \\
 & 
 \bar{\Xi}_i =1 +\ellsquare  \baa_i^2  
 \,.
 &
 \end{align}
 There are various fortunate cancellations, as follows.
 We start by calculating the coefficient of $(d|\bmu|)^2$ in \eqref{19XII24.2}: 
 \begin{align}
 &
 \sum_{i=1}^N\frac{\bar{r}^2+\baa_i^2}{\bar{\Xi}_i} (n^i)^2 d|\bmu|^2
 - 
 \frac {r^2(d|\bmu|)^2}{1 +  |\bmu|^{-2}} 
 - \frac{\ellsquare  (1-\ellsquare \bar{r}^2)}{1 +  \sum_{i=1}^{N }
 	\frac{ \bar a_i^2(n^i)^2|\bmu|^2}{\ell^2 \bar\Xi_{i}} }
 \Bigl(
 \sum_{i=1}^N \frac{\baa_i^2}{\bar{\Xi}_i}(n^i)^2 |\bmu| d|\bmu|
 \Bigr)^2 
 \nn
 \\
 & =
 \sum_{i=1}^N\frac{\bar{r}^2+\baa_i^2}{\bar{\Xi}_i} (n^i)^2 d|\bmu|^2
 - \frac {r^2(d|\bmu|)^2}{1 +  |\bmu|^{-2}} 
 - \frac{  (1-\ellsquare \bar{r}^2)}{\ell^2|\bmu|^{-2} +  \sum_{i=1}^{N }
 	\frac{ \bar a_i^2(n^i)^2}{ \bar\Xi_{i}} }
 \Bigl(
 \sum_{i=1}^N \frac{\baa_i^2 (n^i)^2}{\bar{\Xi}_i}  
 \Bigr)^2 (d|\bmu|)^2
 \nn
 \\
 & =
 \sum_{i=1}^N\frac{\bar{r}^2+\baa_i^2}{\bar{\Xi}_i} (n^i)^2 d|\bmu|^2
 - r^2(d|\bmu|)^2
 - (1-\ellsquare \bar{r}^2)
 \Bigl(
 \sum_{i=1}^N \frac{\baa_i^2 (n^i)^2}{\bar{\Xi}_i}  
 \Bigr) (d|\bmu|)^2 
 \nn
 \\
 &\quad
 +\ell^2 |\bmu|^{-2}(d|\bmu|)^2 
 + O(|\bmu|^{-4})
 \nn
 \\
 & = 
 \underbrace{\sum_{i=1}^N (n^i)^2 r^2 (d|\bmu|)^2 - r^2 (d|\bmu|)^2}_{=0}
 \quad+\ell^2|\bmu|^{-2}(d|\bmu|)^2 
 + O(|\bmu|^{-4})
 \,.
 \end{align}
 Next, the terms involving $dn^i$ and the remaining terms involving  $d|\bmu|$ read
 \begin{align}
 2|\bmu|
 \Big(
 &\sum_{i=1}^N\frac{\bar{r}^2+\baa_i^2}{\bar{\Xi}_i} n^idn^i
 \Big)  d|\bmu|
 + |\bmu|^2\sum_{i=1}^N\frac{\bar{r}^2+\baa_i^2}{\bar{\Xi}_i} (dn^i)^2
 \nn
 \\
 &\   -\frac{\ellsquare  (1-\ellsquare \bar{r}^2)}{\bar{W}}
 \bigg[
 |\bmu|^4
 \big(\sum_{i=1}^{N}\frac{ \bar a_i^2 n^i dn^i}{\bar \Xi_i}\big)^2 
 + 2|\bmu|^3 (\sum_{i=1}^{N}\frac{ \bar a_i^2 (n^i)^2 d|\bmu|}{\bar \Xi_i})(\sum_{j=1}^{N}\frac{ \bar a_j^2 n^j dn^j}{\bar \Xi_i})
 \bigg]
 \nn
 \\
 =&\ 
 2|\bmu|
 \big(\sum_{i=1}^N\frac{\bar{r}^2+\baa_i^2}{\bar{\Xi}_i}n^i dn^i\big)
 d|\bmu|
 + |\bmu|^2\sum_{i=1}^N\frac{\bar{r}^2+\baa_i^2}{\bar{\Xi}_i} (dn^i)^2
 \nn
 \\
 &  -\frac{  (1-\ellsquare \bar{r}^2)|\bmu|^2}{\ell^2|\bmu|^{-2} +  \sum_{i=1}^{N }
 	\frac{ \bar a_i^2(n^i)^2}{ \bar\Xi_{i}} }
 \bigg[
 \Big(\sum_{i=1}^{N}\frac{ \bar a_i^2 n^i dn^i}{\bar \Xi_i}\Big)^2 
 + 2|\bmu|^{-1} 
 \Big(\sum_{i=1}^{N}\frac{ \bar a_i^2 (n^i)^2 d|\bmu|}{\bar \Xi_i}\Big)
 \Big(\sum_{j=1}^{N}\frac{ \bar a_j^2 n^j dn^j}{\bar \Xi_i}\Big)
 \bigg]
 \nn
 \\
 =&\ 
 2|\bmu| \big(\sum_{i=1}^N\frac{\bar{r}^2+\baa_i^2}{\bar{\Xi}_i}n^i dn^i\big)
 d|\bmu|
 + |\bmu|^2\sum_{i=1}^N\frac{\bar{r}^2+\baa_i^2}{\bar{\Xi}_i} (dn^i)^2 
 \nn
 \\
 & 
 -\frac{  (1-\ellsquare \bar{r}^2)|\bmu|^2}{ \sum_{i=1}^{N }
 	\frac{ \bar a_i^2(n^i)^2}{ \bar\Xi_{i}} }
 \Big(\sum_{i=1}^{N}\frac{ \bar a_i^2 n^i dn^i}{\bar \Xi_i}
 \Big)^2 
 -2 (1-\ellsquare \bar{r}^2)|\bmu|
 \Big(\sum_{j=1}^{N}\frac{ \bar a_j^2 n^j dn^j}{\bar \Xi_i}\Big)d|\bmu|
 \nn
 \\
 &\ 
 + O(1)dn^i dn^j +    O( \red{|\bmu|^{-1}}) dn^i d|\bmu|  
 \nn
 \\
 =&\ 
 2|\bmu|
 \Big(r^2 \underbrace{\sum_{i=1}^N n^idn^i}_{=0}
 \Big)
 d|\bmu|
 + |\bmu|^2
 \sum_{i=1}^N\frac{\bar{r}^2+\baa_i^2}{\bar{\Xi}_i} (dn^i)^2 
 -\frac{  (1-\ellsquare \bar{r}^2)|\bmu|^2}{ \sum_{i=1}^{N }
 	\frac{ \bar a_i^2(n^i)^2}{ \bar\Xi_{i}} }
 \Big(\sum_{i=1}^{N}\frac{ \bar a_i^2 n^i dn^i}{\bar \Xi_i}\Big)^2  
 \nn
 \\
 & 
 + O(1)dn^i dn^j +   O( \red{|\bmu|^{-1}}) dn^i d|\bmu|  
 \,.
 \end{align}
 Hence
 \begin{align}\label{19XII24.6}
 \ringh_\KS
 = 
 & \    \ell^2 
 |\bmu|^{-2} 
 \big( 1 +
 O(  |\bmu|^{-2} )
 \big)
 (d|\bmu|)^2 +   O( \red{|\bmu|^{-1}})dn^i d|\bmu|  
 \nonumber
 \\
 & \ +  |\bmu|^{2} 
 \Bigg[   
 \mysigma_{ij} dn^i dn^j
 +
 \sum_{i=1}^N\frac{\bar{r}^2+\baa_i^2}{\bar{\Xi}_i}
 \bigl((n^i)^2 d\red{\varphi^i})^2  
 %\nonumber
 % \\
 % & 
 +\frac{2\bar m\br}{|\bmu|^{-2} \barUh}
 \Big(
 \sum_{i=1}^{N}\frac{ \bar a_i (n^i)^2}{\bar \Xi_i}d\red{\varphi^i}
 \Big)^2
 \Bigg]\,,
 \end{align}
 with
 \begin{equation}\label{1II25.1}
 \mysigma_{ij}dn^i dn^j = 
 \sum_{i=1}^N\frac{\bar{r}^2+\baa_i^2}{\bar{\Xi}_i} (dn^i)^2 
 -\frac{  (1-\ellsquare \bar{r}^2) }{ \sum_{i=1}^{N }
 	\frac{ \bar a_i^2(n^i)^2}{ \bar\Xi_{i}} }
 \Big(\sum_{i=1}^{N}\frac{ \bar a_i^2 n^i dn^i}{\bar \Xi_i}\Big)^2  
 + O(|\bmu|^{-2})dn^i dn^j 
 \,.
 \end{equation}
 In order to  check that the limit as $|\mu|$ tends to $\infty$ of the tensor field  $ \mysigma$ is  non-degenerate,  we set
 \begin{equation}\label{2I25.12}
 \theta^i = \sqrt{\frac{\bar{r}^2+\baa_i^2}{\bar{\Xi}_i}} dn^i
 \,,
 \qquad
 Z^i = 
 \sqrt{\frac{\bar \Xi^i   (1-\ellsquare \bar{r}^2) }{ (\bar{r}^2+\baa_i^2)
 		\sum_{j=1}^{N } 
 		\frac{ \bar a_j^2(n^j)^2}{ \bar\Xi_{j}} }
 }
 \frac{ \bar a_i^2 n^i }{\bar \Xi_i}
 \,.
 \end{equation}
 Given a vector field $X$, with  $X^i:=\theta^i(X)$, we have
 \begin{equation}\label{2I25.13}
 \mysigma(X,X) =\sum_{i=1}^{N} (X^i)^2 - (\sum_{i=1}^{N} X^i Z^i)^2
 \,.
 \end{equation}
 This will be positive for all non-zero vectors $X^i$ if 
 \begin{equation}\label{2I25.14}
 \sum_{i=1}^{N} (Z^i)^2  < 1
 \,.
 \end{equation}
 Equivalently,
 \begin{equation}\label{2I25.15}
 \sum_{i=1}^{N}   \frac{    (1- \ellsquare\bar{r}^2)\bar a_i^4 (n^i)^2 }{ \bar \Xi_i(\bar{r}^2+\baa_i^2)
 } 
 < \sum_{j=1}^{N } 
 \frac{ \bar a_j^2(n^j)^2}{ \bar\Xi_{j}}
 \,,
 \end{equation}
 which is the same as
 \begin{equation}\label{2I25.16}
 \sum_{i=1}^{N}   \frac{    (\ell ^2- \bar{r}^2)\bar a_i^4 (n^i)^2 }
 { (\ell ^2 +  \baa_i^2 )(\bar{r}^2+\baa_i^2)
 } 
 < \sum_{j=1}^{N } 
 \frac{ \ell^2 \bar a_j^2(n^j)^2}{ \ell ^2 +  \baa_j^2 }
 \,,
 \end{equation}
 or as
 \begin{equation}\label{2I25.17}
 \sum_{i=1}^{N}   \frac{    
 	\big((\cancel{\ell ^2}- \bar{r}^2)\bar a_i^2 -(\bar{r}^2+\cancel{\baa_i^2}) \ell^2
 	\big)  \baa_i^2   (n^i)^2 }
 { (\ell ^2 +  \baa_i^2 )(\bar{r}^2+\baa_i^2)
 } 
 <  0 
 \,,
 \end{equation}
 which is obviously satisfied.

 Recall that $|\bmu|^{-2} \barUh$ has a smooth limit as $|\bmu|$ tends to infinity, with the limit bounded away from zero. Replacing $|\bmu|$ by  
 $ 1/x$ we find 
 \begin{align}\label{19XII24.7}
 \ringh_\KS
 = 
 & \   
 x^{-2} 
 \Bigg[
 \big(  \ell^2  +
 O(  x^{ 2} )
 \big)
 dx^2 +   O(x)dn^i dx 
 \nonumber
 \\
 & \ +    
 \mysigma_{ij} dn^i dn^j
 +
 \sum_{i=1}^N\frac{\bar{r}^2+\baa_i^2}{\bar{\Xi}_i}
 \big((n^i)^2 d\red{\varphi^i}\big)^2  
 %\nonumber
 % \\
 % & 
 +\frac{2\bar m\br}{x^{2} \barUh}
 \Big(
 \sum_{i=1}^{N}\frac{ \bar a_i (n^i)^2}{\bar \Xi_i}d\red{\varphi^i}
 \Big)^2
 \Bigg]\,,
 \end{align}
 which shows that $	\ringh_\KS$ is conformally smooth.
 It is a standard fact that the Riemann tensor of such metrics approaches a negatively curved space form as $|\bmu|$ tends to infinity; equivalently, $x$ tends to zero. This implies in particular that, for any given  $\bm$ and $\br$, the metric   $	\ringh_\KS$ is negatively Ricci curved for sufficiently large $|\bmu|$ and for all rotation parameters, and that it is negatively Ricci curved everywhere for sufficiently small rotation parameters.

 \subsubsection{Topology}
 
 Suppose that there is a quotient map which a) preserves the Killing vector $\partial_\tau$, and which b)  compactifies the horizon, and under which c) the whole spacetime metric passes to the quotient. 
 Then in particular the one-form
 \begin{equation}\label{7XII24.4}
 \red{\alpha} : =  
 \bg(\partial_\tau,\cdot)
 \equiv   \bg(\partial_\tau,\partial_{x^\mu}) dx^\mu
 = 
 \bg(\partial_\tau,\partial_{\varphi^i} ) d\varphi^i 
 = \bar W 
 \Big(
 \sum_{i=1}^{N}\frac{ \bar a_i\bmuisq}{\bar \Xi_i}d\red{\varphi^i}
 \Big)
 \end{equation}
 will also pass to the quotient.
 But we have:
 
 \begin{Proposition}
 	\label{P4XII24.1b} 
 	Let $\br_H$ be a solution of $ \bar V(\br_H) = 2 \bm \br_H$. 
 	If $(\red{\mcN_{\btau,\br_H}},\ringh_\KS,\red{\alpha})$ can  be compactified with  a negative-definite Ricci tensor of $\ringh$, then all rotation parameters vanish. 
 \end{Proposition} 

 \proof
 Set  
 \begin{equation}\label{10XII24.1a}
 X:=\ringh^{ij}_\KS \red{\alpha}_j\partial_{\red{\varphi^i}} 
 \,.
 \end{equation}
 By an abuse of notation we will use the same symbols as in \eqref{15XI24.4}-\eqref{16XI24.91} for the corresponding formulae in Kerr-Schild coordinates. 
 Then, as $h^{ij}$ is the $\varphi^i$-block in $\ringh^{ij}_\KS$,  we find 
 \begin{equation}\label{10XII24.1b}
 X= h^{ij}  \red{\alpha}_j\partial_{\red{\varphi^i}}=
 N^i \partial_{\red{\varphi^i}}
 =  
 \underbrace{ 2\bm\br \bW
 	\left({\barUh}+{2\bm\br}\sum_{i=1}^N\frac{\baa_i^2\bmuisq}{(\br^2+\baa_i^2)\bXi_i} \right)^{-1}}_{(\star)}   
 \sum_{i=1}^N\frac{\baa_i}{\br^2+\baa_i^2}
 \partial_{\red{\varphi^i}}
 \,.
 \end{equation}
 The same calculation as in \eqref{28VII24.41}-\eqref{13XII24f.1} gives
 \begin{align}
 (*)^{-1}|_{\br=\br_H} &=
 \frac{\ell^2}{(\br^2-\ell^2)\bW2\bm\br}\left((V-2\bm\br)\left[1+\sum_{i=1}^N \frac{\baa_i^2}{\br^2+\baa_i^2}\right]+2\bm\br\bW\right)\Bigg|_{\br=\br_H}
 \nonumber
 \\
 &=\frac{\ell^2}{\br_H^2-\ell^2}\,.
 \end{align}
 Thus $ (*)|_{\br=\br_H} $ is constant on each horizon.   
 
 Since $\bar\alpha$ passes to the quotient, the vector field $X$ is well defined on the quotient, tangent to $\red{\mcN_{\tau,\br_H}}$, and is a linear combination of $\partial_{\red{\phi^i}}$'s with constant coefficients there, hence a Killing vector of $(\red{\mcN_{\tau,\br_H}},\ringh_\KS)$. 
 But compact Ricci-negatively-curved manifolds have only trivial Killing vectors, thus $X=0$; equivalently, all the rotation parameters vanish. 
 \qedskip

 It follows that the cross-sections of the horizon cannot be compactified in the slowly-fully-rotating case, since on a compact manifold and for small rotation parameters the tensor field  $\ringh_\KS$  would be  Ricci-negatively-curved.  
 
 Our results are complemented by the statement  in the second paragraph of page three in \cite{Klemm:1998kd} (correcting the original work~\cite{Klemm:1997ea}),   that both the four-dimensional  and higher-dimensional solutions with \emph{exactly one} non-zero rotation parameter, whether small or large, cannot be compactified without creating a discontinuity in the metric.  
 
 Compare~\cite{peraza2024staticvacuum31black} for an apparently similar behaviour.

 \subsection{Black hole regions}
 \label{s14XI24.1}
 
 We consider the manifold
 \begin{equation}\label{14XII24.2a}
 \mcM_{\mathrm{KS}}:=\{ \br\in \R\}\times \R^{2N+1}
 \,.
 \end{equation}
 as defined at the end of Section~\ref{ss28VII24.1}, with a globally defined function $\br$ ranging over $\R$. 
 By an abuse of notation we will continue to denote by $\bg$ the metric on  $\mcM_{\mathrm{KS}}$. 
 For definiteness we assume a fully-rotating configuration.

 We claim that $(\mcM_{\mathrm{KS}},\bar g)$  contains a black hole region or a white hole region, depending upon the choice of time orientation. 
 To see this, recall that the gradient of  $\bar r$ is a time function on 
 the set $\{\barVh - 2 \bm \br < 0\}\cap \{  \br <\br_H\}$.
 Let us choose the time orientation so that $-\br$ is a time function on $\{\barVh - 2 \bm \br < 0\}$ near $\br_H$, thus $\bar r $ is  strictly decreasing along every causal curve in this last region. Hence no causal curve that intersects   
 $$
 \mcB:= \{ \br < \bar r_H\}
 $$
 can leave this region across $\{\br=\br_H\} $.
 If $\scri$ denotes the conformal completion at infinity constructed in Section~\ref{ss28VII24.4}, we conclude that 
 $$
 J^+( \mcB) \cap J^-(\scri) \subset J^+( \mcB) \cap J^-(\{\br>\br_H\})  
 = \emptyset\,,
 $$
 which is the definition of a black hole region.

\section{Projection diagrams}
\label{s15XI24.1}

Projection diagrams~\cite{ChBlackHoles,COS} provide a generalisation of the Carter-Penrose diagrams to spacetimes which are not necessarily two-dimensional or  spherically symmetric.  
They provide a convenient way to visualise the global causal structure of extensions. 
The idea is to construct a projection from a spacetime  $(\mcM,\bar g)$ to a subset, say $\mcU$,  of two-dimensional Minkowski spacetime $\R^{1,1}\equiv (\R^2,\eta)$ so that 
\begin{enumerate}
	\item $\bar g$-causal curves project to $\eta$-causal curves, and  
	\item every $\eta$-timelike curve in $\mcU$ is the projection of some $\bar g$-timelike curve in  $(\mcM,\bar g)$.
\end{enumerate}

In order to construct such  diagrams for the extensions of the metric \eqref{3VI24.f78},  the simplest approach is to project-out the coordinates $\red{\bmu^i}$ and $\red{\phi^i}$, so that we are left with two local coordinates $\bar t$ and $\bar r$. 
Now, it follows from the calculations in Appendix~\ref{s5XI24.1} below that the $\red{\bmu^i}$-sector of the metric is positive definite. 
So we start by projecting-out the $\red{\bmu^i}$-coordinates and  consider the tensor field  
\begin{align}\label{15XI24.2}
\newg  :=\ &
\bar{W}(1-\ellsquare\bar{r}^2)d\bar{t}^2
+\frac{\barUh d\bar{r}^2}{\barVh-2\bar{m}\br}
+\frac{2\bar{m}\br}{\barUh}
\biggl(\bar{W}d\bar{t} 
+  \sum_{i=1}^N \frac{\baa_i \bmuisq d\red{\phi^i}}{\bar{\Xi}_i}   
\biggr)^2
\nonumber
\\
& 
+\sum_{i=1}^N
\frac{\bar{r}^2+\baa_i^2}{\bar{\Xi}_i} 
\bmuisq\red{(d\phi^i)^2} 
\,,
\end{align}
which has  signature $(1,N+1)$ and satisfies 
\begin{equation}\label{16XI24.5}
\forall \ X \qquad 
\bar g(X,X) \ge \newg(X,X)   
\,,
\end{equation}
with equality if and only if $X$ has vanishing $\red{\bmu^i}$-components. As the neglected part of $\bg$ is positive definite, all $\bg$-causal curves $\gamma$ are $\newg$-causal:
$$
\bg(\dot{\gamma}, \dot{\gamma})\le 0 
\quad
\Longrightarrow
\quad
\newg(\dot{\gamma}, \dot{\gamma})\le \bg(\dot{\gamma}, \dot{\gamma})\le 0 
\,.
$$
Moreover, if  the coordinates $\red{\bmu^i}$ are constant along a curve $\gamma$, then   $\gamma$ will be causal for $\newg$ if and only if it is causal for $\bg$.

Recall that $\hat{g}$ denotes the tensor field  induced by $\newg$ on the level sets of $\br$.
Its ADM-decomposition  with respect to the level sets of $\bar t$ can be found in \eqref{15XI24.4}-\eqref{16XI24.91}. 
Let $\bgamma$ denote the symmetric tensor field  
\begin{equation}\label{16XI24.2}
\bgamma :=  \pm N^2 d \bar t^2  
+\frac{\barUh d \bar r^2 }{\barVh-2\bar{m}\br}
\,. 
\end{equation}
with $\pm N^2$ given by \eqref{16XI24.91}, 
thus $\bgamma$ is obtained by ignoring the $h$-part of $\hat g$.
It follows from \eqref{16XI24.5} and \eqref{15XI24.4}   that, \emph{in the region where $h$ is positive definite}  we have,
for any vector $X$, 
\begin{equation}\label{16XI24.1}
\bgamma(X,X) =  \pm N^2 (X^{\bar t})^2  
+\frac{\barUh  }{\barVh-2\bar{m}\br}  (X^{\bar r})^2
\le  \newg(X,X)   \le  \bar g(X,X)  
\,.
\end{equation}
As before, this implies that $\bg$-causal curves project to $\bgamma$-causal ones. Furthermore, a $\bgamma$-causal curve $s\mapsto \big(\bt(s),\br(s)\big)$ can be lifted to a $\bg$-causal curve $\gamma(s)$ in spacetime by choosing any constants $\red{\bmu^i} $ and setting
\begin{equation}\label{17XI24.1}
\gamma(s) = \big(\bt(s),\br(s), \red{\bmu^i}, \red{\phi^i}(s)\big)
\,,
\quad
\mbox{where}
\quad
\frac{d\red{\phi^i}}{ds} = - N^i \frac{d\bt}{ds}
\,.
\end{equation}
In order to avoid analysing what happens with lifts passing through $\red{\bmu^i}=0$,  
where the projection map has the standard singularity associated with axes of rotation, 
it is simplest to choose the lift so that all the $\red{\bmu^i}$'s are different from zero, which we do.  

The next step is  to simplify $\bgamma$ using a conformal transformation. For this we decompose $\bUh$ as 
\begin{equation}\label{16XI24.6}
\bUh    =  \underbrace{\bigg(1
	+ \sum_{i=1}^N \frac{ {\bmuisq}\baa_i^2}
	{\br^2+\baa_i^2}\bigg)
}_{
	\brU } \prod_{i=1}^N (\bar{r}^2+\baa_i^2) =:
\brU  \myF ^{2}
\end{equation}
and write 
\begin{equation}\label{16XI24.7}
\bgamma  =\brU 
\Big(
\pm  \breve U^{-1}N ^2 d \bar t^2  
+\frac{\myF ^2 }{\barVh-2\bar{m}\br} d \bar r^2  
\Big)
=: \brU \zgamma
\,,
\end{equation}
where the conformal factor has been chosen so that $ \brU^{-1} \bgamma_{\br\br}$ depends only upon $\br$.  

Note that causality with respect to $\bgamma$ is the same as that for $\zgamma$.

In order to extract out of $\zgamma \equiv \brU^{-1} \bgamma$  a two-dimensional Lorentzian metric which depends only upon $\br$,
we define
\begin{equation}\label{26XI24.1}
\myH_-(\br):=\inf_{\red{\bmu^i}\in \R^N } \frac{ \breve U^{-1}N ^2 }{
	|\barVh-2\bar{m}\br|}
\ \mbox{and} 
\
\myH_+(\br):= \sup_{\red{\bmu^i}\in \R^N } \  \frac{ \breve U^{-1}N ^2 }{|\barVh-2\bar{m}\br|}
\,.
\end{equation}
We show in Appendix~\ref{s17XI24.1} that the functions $ \myH_\pm$ are locally Lipschitz, with $\myH_->0$.  

\subsection{$\barVh-2\bar{m}\br <0$}  

On this region  the coordinate $\br$ has timelike gradient, which implies that
the matrix  $h$ of \eqref{14XII24.3}  is positive definite. 
Recall that
\begin{align}\label{16XI24.3nn} 
\pm N^2 = \ & -\frac{\bW (\barVh-2\bm\br)}{\left(1+\frac{2\bm\br}{\barUh}
	\sum_{i=1}^N\frac{\baa_i^2\bmuisq}{(\br^2+\baa_i^2)\bXi_i}\right)\prod_{i=1}^{N}(\br^2+\baa_i^2)}
\,,
\end{align} 
and since $\bW \ge 1$ we need to choose the positive sign in the  left-hand side of \eqref{16XI24.3nn}.
Then
\begin{align}\label{17XI24.3}
\zgamma(X,X)
= \ &
\breve U^{-1}N ^2 (X^{\bar t})^2 
+\frac{\myF ^2 }{\barVh-2\bar{m}\br} (X^{\br})^2 
\nonumber
\\   
\ge \ & 
|\barVh-2\bar{m}\br|  \myH_- (X^{\bar t})^2 
- \frac{\myF ^2 }{|\barVh-2\bar{m}\br|} (X^{\br})^2 
\nonumber
\\
=:  \ & 
\gamma_-(X,X)
\,,
\end{align}
with equality attained at points where $\myH_-$ is attained.
Note that a vector which is $\zgamma$-causal will also be $\gamma_-$-causal.

From what has been said  the metric $\gamma_-$ has the desired causality properties with respect to the metric $\bg$, and provides the desired \emph{projection metric} with $\bt$ ranging over $\R$ and $\br$ ranging over a connected component of the set
$\{\barVh-2\bar{m}\br <0\}$.

\subsection{$\barVh-2\bar{m}\br >0$} 

On this region  the coordinate $\br$ has spacelike gradient, which implies that the tensor field $\gamma$ of \eqref{15XI24.4} has  Lorentzian signature.
On the subset where the matrix  $h$ of \eqref{14XII24.3} is positive definite we have $\pm N^2 <0$, hence we need to choose the negative sign in the left-hand side of \eqref{16XI24.3nn}. Then
\begin{align}\label{17XI24.4}
\zgamma(X,X)
= \ &
- \breve U^{-1}N ^2 (X^{\bar t})^2 
+\frac{\myF ^2 }{\barVh-2\bar{m}\br} (X^{\br})^2 
\nonumber
\\   
\ge \ & 
- (\barVh-2\bar{m}\br)\myH_+(X^{\bar t})^2 
+  \frac{\myF ^2 }{\barVh-2\bar{m}\br} (X^{\br})^2 
\nonumber
\\
=:  \ & 
\gamma_+(X,X)
\,,
\end{align}
with equality attained at points where $\myH_+$ is attained.
As before we conclude that the metric $\gamma_+$ has the desired causality properties with respect to the metric $\bg$, and provides the desired projection metric on each connected component of the set
\begin{equation}\label{17XI24.5}
\{\barVh-2\bar{m}\br >0\} \cap \{\mbox{the matrix $h(\partial_{\red{\phi^i}},\partial_{\red{\phi^j}})$ is positive definite}
\}
\,.
\end{equation}

We emphasise that no two-dimensional metric can exist with the properties listed at the beginning of this section in the region where $h$ is not positive definite, as there are closed timelike curves throughout the interior of that region, as well as closed causal curves on its boundary.  

\subsection{Further extensions}

\begin{figure}[ht]
	\centering
	\includegraphics[scale=.8]{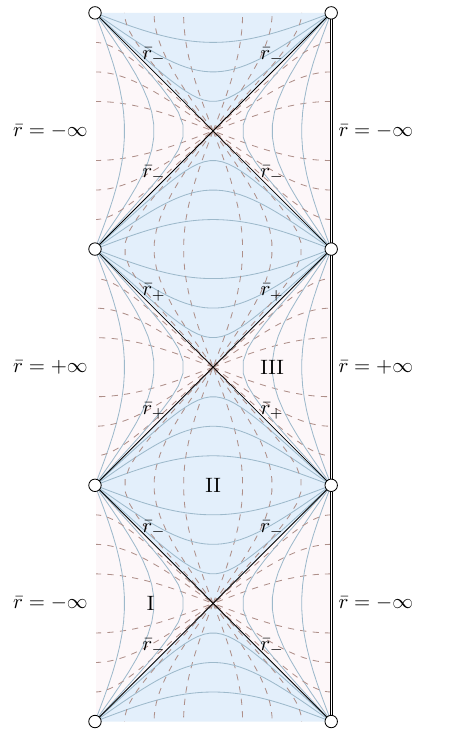}
	\caption{The projection diagram for an extended spacetime with $\bm>0$, in which there are no closed causal curves, with $\bar V - 2 \bm \br$  having two zeros $-\ell <\br_- <\ell < \br_+$. 
		Both horizons are of bifurcate-type, with   non-zero surface gravity.  The light blue regions are where $\nabla\br$ is timelike and the pink ones where $\nabla\bt$ is timelike.  
		The solid-blue curves represent the level sets of constant $\br$ while the dashed-red those of constant $\bt$. 
		The Kerr--Schild extension is indicated by the regions marked I-III.
		The diagram can be continued indefinitely in the vertical direction so that a  shift of the figure by two diamonds generates a $\Z$-group of isometries. Distinct maximal analytic extensions can be obtained by quotienting the spacetime by distinct subgroups of $\Z$, leading to closed timelike curves through every point. Further  distinct maximal analytic extensions can be obtained by removing, e.g., one or more bifurcation surfaces and passing to covering spaces. 
	}\label{F17IX24.1}
\end{figure}

\begin{figure}
	\centering
	\includegraphics[width=.7\textwidth]{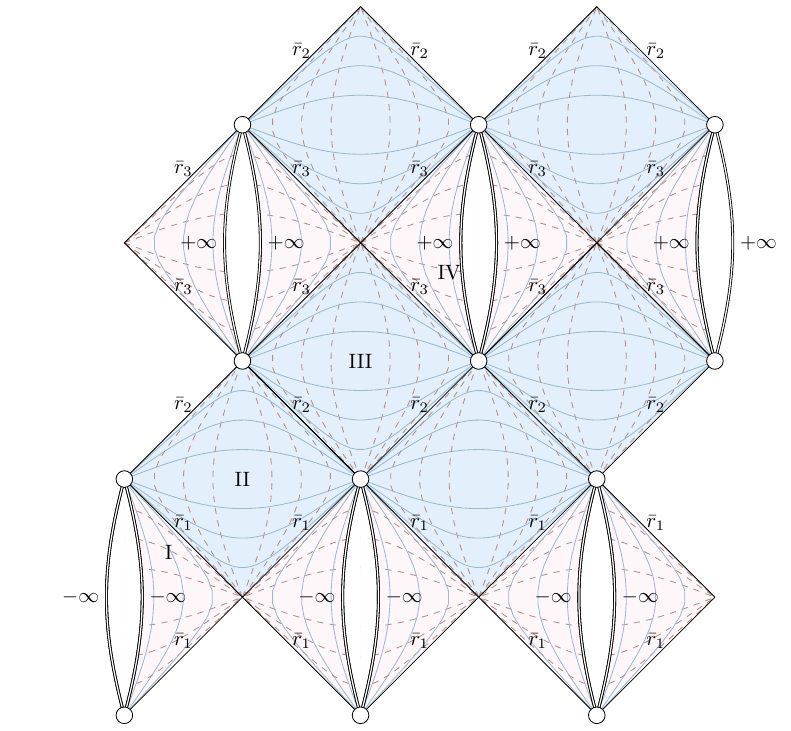}
	\caption{The projection diagram for an extended spacetime with $\bm>0$   in which there are no closed causal curves and $\bar V - 2 \bm \br$ has three zeros $-\ell <\br_1 < \br_2 <\ell < \br_3$,  
		with the zero at $\br_2$ of order two.  
		The coloring of diamonds and the dotting of curves
		are as in Figure~\ref{F17IX24.1}. The extension can be continued in all directions.
		The horizons at $\br_2$ are  degenerate, in the sense that they have vanishing surface gravity $\kappa$, the other ones have non-zero $\kappa$.
	}\label{F17IX24.2}
\end{figure}

\begin{figure}
	\centering
	\includegraphics[width=.8\textwidth]{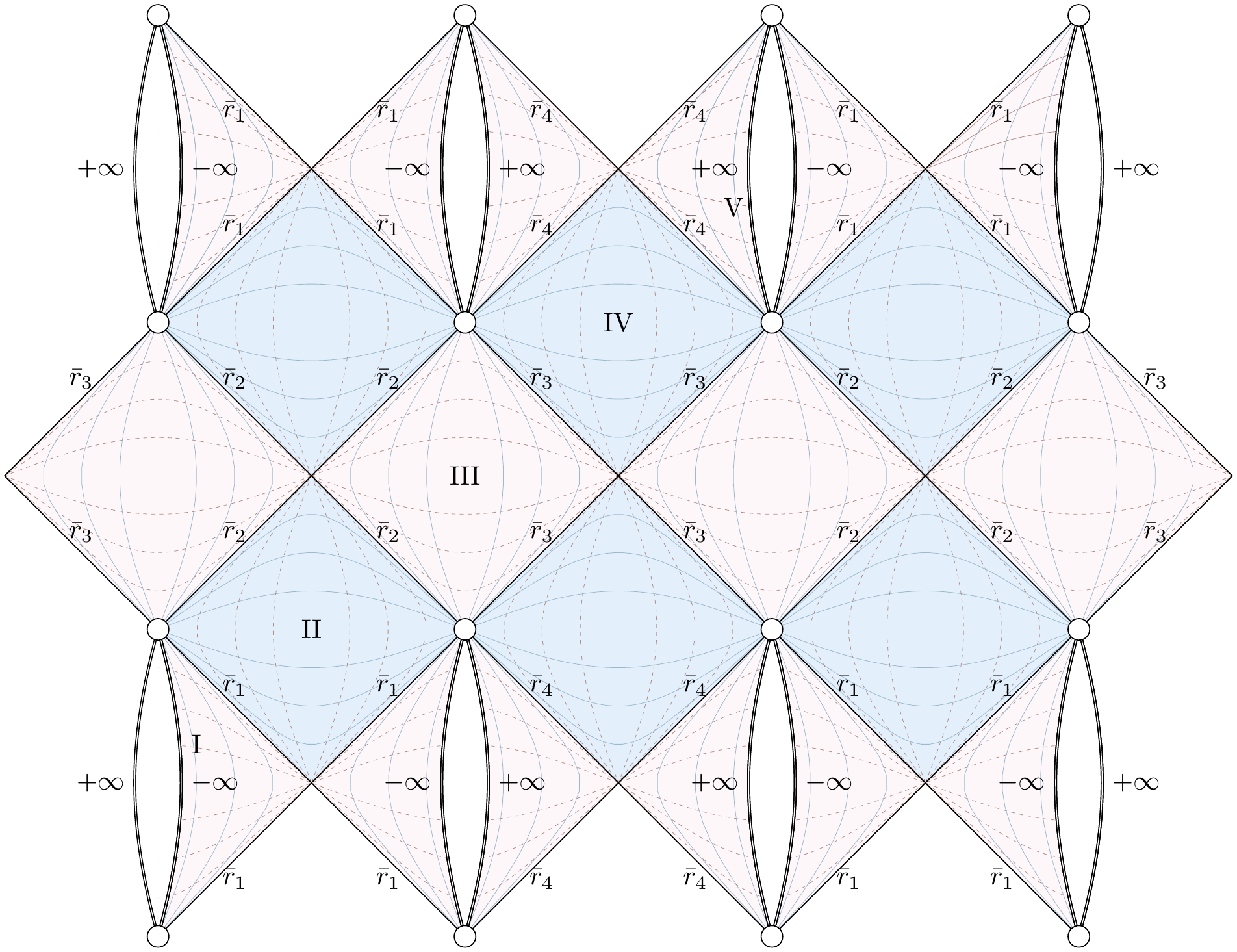}
	\caption{The projection diagram for an extended spacetime with $\bm>0$   in which there are no closed causal curves and $\bar V - 2 \bm \br$ has four  zeros $-\ell< \br_1 < \br_2< \br_3 <\ell < \br_4$.   The extension can be continued to the whole Minkowskian plane in all directions.
		All horizons are of bifurcate-type, with   non-zero surface gravity. 
		The nature of the diamonds and of the curves are as in Figure~\ref{F17IX24.1}. 
		The Kerr--Schild extension is the union of  the regions marked I to V.
		The extension is symmetric with respect to reflections across vertical and horizontal lines passing through the bifurcation points of the Killing horizons.
		Shifts of the plane by two diamonds along the vertical axis and by two diamonds along the horizontal axis generate a $\Z^2$ group of isometries of the depicted spacetime when extended to the whole of $\R^2$;  taking quotients by  distinct subgroups of $\Z^2$ leads to distinct, non-isometric, analytic extensions of the original metric, possibly but not necessarily introducing causality violations.
		Further distinct extensions can be obtained by passing to distinct coverings of the projection plane.
	}  
	\label{F17IX24.3}
\end{figure}

From what has been said above, and as explained in~\cite{ChBlackHoles}, the Kerr-Schild extension of $\bg$ describes a spacetime the global structure of which can be visualised by the numbered regions of the projection diagrams of Figures~\ref{F17IX24.1}-\ref{F17IX24.3}. 
These diagrams are relevant for parameters $\bm,\baa_i$   such that there are no causality violating regions.

The extension to the non-numbered regions of the figures  can be carried-out by the methods explained in~\cite{RaczWald1}.
Whether or not the spacetimes visualised in these diagrams are maximal  requires an analysis of extendibility of the metric as $|\bmu|$ tends to infinity, which we have not carried out. 


\begin{figure}
	\begin{center}
		\includegraphics[scale=0.8,angle=0]{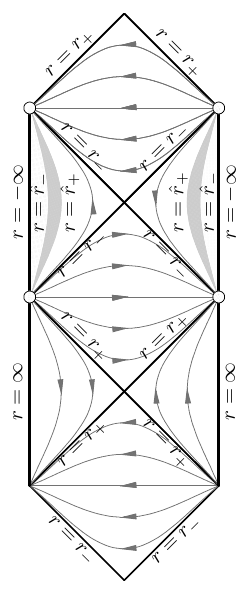}
		\hspace{1cm}
		\includegraphics[scale=0.8,angle=0]{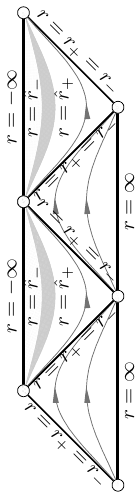}
		\caption{
			Projection diagrams for the Kerr--Newman anti-de Sitter metrics with two distinct zeros of the Boyer-Lindquist function $\Delta_r$ (left diagram) and one double zero (right diagram), from~\cite{COS}, generated using~\cite{SzybkaOelz}. The arrows represent the orientation of the orbits of the Killing vector $\partial_t$.  The shaded area is the projection of the Carter time-machine region, including the singular ring, and should be removed from the diagram for a faithful representation of causality.   An identical diagram is obtained for our metrics in the two-zeros case of Figure~\ref{F17IX24.1} when causality violations occur with $\bm>0$.
			\label{F27XI24.1}}
	\end{center}
\end{figure}

For comparison we present the projection diagrams for  the four-dimensional spherical Kerr-Newman anti-de Sitter metrics in  Figure~\ref{F27XI24.1}.
Identical diagrams are obtained as the one on the left in our  negatively-curved higher dimensional case when causality violations occur in the two-zeros case. 
In the spherical four-dimensional case the excluded shaded region is the projection of both the singular ring and the causality-violating region, while in the negatively curved case only causality violations occur.
\section{Odd dimensions? Positive $\Lambda$?}
\label{App24XI24.1}

One might seek to generalise our analysis to odd spacetime dimensions. 
It turns out that the simplest generalisation of the scheme \eqref{3VI24.f6} to odd dimensions, i.e.
\begin{align}
\mu_j&= i \bmu_j \,,\quad 1\leq j<N-1\,,\qquad \text{and}\qquad
\mu_{N}=\bmu_{N}
\nonumber\\
t&=i\bar{t}\,,\quad r= -i\bar{r}\,, \quad \, m=i^{-(d-1)}\bar{m}\,, \quad a_i=-i\baa_i\,,\quad 1\leq i<N\,,\qquad
\end{align}
does not work. 
The reason for this is, that in odd dimensions the number of rotation parameters $\baa_i$ and angular direction $\red{\phi^i}$ is equal to the number of unconstrained $\red{\bmu^i}$ variables. 
The fact that the last $\bmu_N$ is not scaled in the same way as the rest then causes problems with the signature of the metric. 

We have checked that rescaling all of the $\red{\bmu^i}$'s in the same way and adjusting the constraint does not yield a sensible new metric either.
This appears to be the reason why the previous literature~\cite{Klemm:1997ea,Klemm:1998kd} has not been generalised to all dimensions.

We also remark that our attempts to obtain toroidal cross-sections of the horizon by scaling and passing to the limit did not produce sensible Lorentzian metrics either. 

Last but not least, the metrics obtained by the methods above with a positive cosmological constant are plagued by singularities which we were not been able to resolve in a satisfactory manner, and therefore we do not report on them here.

\bigskip
{\noindent
	\sc
	Acknowledgements:}

PTC's research was supported in part by the NSF under Grant No. DMS-1928930
while he was in residence at the Simons Laufer Mathematical Sciences
Institute (formerly MSRI) in Berkeley during the Fall 2024 semester. 

\appendix

\section{Determinant}
\label{s5XI24.1}

As a step towards determining the signature of $\bar g$ we calculate its determinant.
For this we  invoke the \emph{matrix determinant lemma}.
That is, given an  invertible $m\times m$ square matrix, two $m$-dimensional vectors $u,v$ and some real number $\alpha$, we have the following expression for the determinant of the matrix $M=A+\alpha u v^T$,
\begin{equation}\label{24VIIf.2}
\det(A+uv^T)=(1+\alpha v^T\cdot A^{-1}\cdot u)\det(A)\,.
\end{equation}

We will also make use of the \emph{Sherman–Morrison formula}, which gives an expression for the inverse of $M$. 
That is,
\begin{equation}\label{24VIIf.3}
(A+uv^T)^{-1}=A^{-1}- \alpha \frac{(A^{-1}\cdot u) (v^T\cdot A^{-1})  }{(1+\alpha v^T\cdot A^{-1}\cdot u)}\,,
\end{equation}
which will be useful for calculating the inverse metric.

Since our metric \eqref{3VI24.f7} is block diagonal in three sectors (the $\br$, $x^{\alpha}=(\bar{t},{\phi}_i)$ and $x^{i}=\red{\bmu^i}$ sectors) we can apply these to each in order to calculate the determinant 
\begin{equation}\label{25VII24f.6}
\det (\bar g_{\mu\nu})=\bar g_{\br\br} \det( \hat{g}_{\alpha\beta})\det(\tilde g_{i j})\,.
\end{equation}
Since
$$
\bar g_{\br \br} = \frac{\barUh }{\barVh-2\bar{m}\br}
=  \frac 1 {\barVh-2\bar{m}\br} \bigg(1
+ \sum_{i=1}^N \frac{ {\bmuisq}\baa_i^2}
{\br^2+\baa_i^2}\bigg) \prod_{i=1}^N (\bar{r}^2+\baa_i^2) 
\,,
$$
we see that 
\begin{eqnarray} 
&& \mbox{
	the first multiplicative factor in \eqref{25VII24f.6}  has constant sign
}
\nonumber
\\
&&
\mbox{ wherever $ \barVh-2\bar{m}\br$ has constant sign.
}
\label{6XI24.1}
\end{eqnarray}

Next, let us consider the tensor field $\hat{g}_{\alpha\beta}$ which describes the $(\bt,\red{\phi^i})$-sector of $\bar g$. 
It can be written in the form \eqref{24VIIf.2} where the vectors $u$ and $v$ are collinear, while the matrix $A$ is given by the remaining, diagonal, $(\bt,\red{\phi^i})$ metric components, so the calculation of $A^{-1}$ is immediate.
We find, for $\br^2 \ne \ell^2$,
\begin{align}\label{25VII24f.8}
\det(\hat g_{\alpha\beta})&=\left(1
+\frac{2\bar{m}\br}{\barUh}
\left[\frac{\bar{W}}{1-\ellsquare \bar{r}^2}
+\sum_{i=1}^{N}
\frac{\muisq}{\bar{\Xi}_i}\frac{\baa_i^2}{\bar{r}^2+\baa_i^2}
\right]\right)
% 		 \nonumber
% \\
% 		 &\times 
\bar{W}(1 -\ellsquare \bar{r}^2)\prod_{i=1}^{N}\frac{\muisq}{\bar{\Xi}_i}
(\bar{r}^2+\baa_i^2)
\nonumber\\
&=
\left(1
+
\frac{2\bar{m} \bar{r}}{(1 -\ellsquare \bar{r}^2)
	\prod_{i=1}^{N}(\bar{r}^2+\baa_i^2)} \right)
\bar{W}(1 -\ellsquare \bar{r}^2)\prod_{i=1}^{N}\frac{\muisq}{\bar{\Xi}_i}(\bar{r}^2+\baa_i^2)
\nonumber
\\
&=-
\bar{W}\left(\barVh-2\bar{m}\bar{r}\right)\prod_{i=1}^{N}\frac{\bmuisq}{\bar{\Xi}_i}\,.
\end{align}

While the final result is insensitive to  zeros of  $1-\ellsquare \br^2$, this calculation does not apply when $\br = \pm \ell$. 
A calculation that avoids this problem proceeds by performing 
first the ADM-decomposition
\begin{align}
\hat{g} :=\ &
\bar{W}(1-\ellsquare\bar{r}^2)d\bar{t}^2 
+\frac{2\bar{m}\br}{\barUh}
\biggl(\bar{W}d\bar{t} 
+  \sum_{i=1}^N \frac{\baa_i \bmuisq d\red{\phi^i}}{\bar{\Xi}_i}   
\biggr)^2 
+\sum_{i=1}^N
\frac{\bar{r}^2+\baa_i^2}{\bar{\Xi}_i} 
\bmuisq\red{(d\phi^i)^2} 
\nonumber
\\       
=: \
& 
\pm N^2 d\bar t^2 + h_{ij}(d\red{\phi^i} + N^i d\bar t) (d\red{\phi^j}+ N^j d\bar t)
\,,\label{20XI24.f1}
\end{align}
with
\begin{align}
h \equiv h_{ij} d\red{\phi^i} d\red{\phi^j} 
= \ & 
\underbrace{\sum_{i=1}^N
	\frac{\bar{r}^2+\baa_i^2}{\bar{\Xi}_i} 
	\bmuisq\red{(d\phi^i)^2} }_{A}
+ \frac{2\bar{m}\br}{\barUh}
\underbrace{\biggl( \sum_{i=1}^N \frac{\baa_i \bmuisq d\red{\phi^i}}{\bar{\Xi}_i}   
	\biggr)^2 }_{u u^T}
\,, 
\label{20XI24.f2}
\\
N^j = \ & h^{ij} 
\frac{2\bar m \bar r \bar W\baa_i \bmuisq }{\bar U \bXi_i}
\,, \label{20XI24.f3}
\\
\pm N^2 = \ & \bar{W}(1-\ellsquare\bar{r}^2) 
+ \frac{2\bar{m}\br}{\barUh} \bar{W}^2
-
% \sum_{i,j =1}^N 
h_{ij}N^i N^j
\,.
\label{20XI24.f4}
\end{align}
Next, there is the standard decomposition of the determinant of $\hat{g}$ in terms of $\pm N^2$ and $h$;
\begin{equation}
\det \hat{g}=\pm  N^2 \det h\,.
\label{30XI24.11}
\end{equation}
So we need to calculate $N^2$ and $\det h$. 
The latter is straightforward since $h$ is in a form where we can apply \eqref{24VIIf.2} and \eqref{24VIIf.3}, and in particular, $A$ is diagonal. Therefore
\begin{equation}
\label{25XI24.1}
\det h= \Bigg(1+\frac{2\bm\br}{\barUh} \underbrace{\sum_{i=1}^N\frac{\baa_i^2\bmuisq}{(\br^2+\baa_i^2)\bXi_i}}_{u^TA^{-1}u}\Bigg)\prod_{i=1}^{N}\frac{(\br^2+\baa_i^2)\bmuisq}{\bXi_i}\,.
\end{equation}
We leave the term in brackets as is for now because, using hindsight, it will cancel out of the final expression.

Now to calculate $\pm N^2$ we need to know $N^i$ and hence we need also $h^{ij}$. 
Assuming from now on that we are  away from zeros of 
$
\left(1+\frac{2\bm\br}{\barUh}\,u^TA^{-1}u\right)$,
using \eqref{24VIIf.3} we obtain
\begin{align}
h^{i j}=\frac{\bXi_i}{(\br^2+\baa_i^2)\bmuisq}\delta^{ij}- \frac{2\bm\br}{\barUh
	\left(1+\frac{2\bm\br}{\barUh}\,u^TA^{-1}u\right)
}\frac{\baa_i\baa_j}{(\br^2+\baa_i^2)(\br^2+\baa_j^2)}\,.
\end{align}
This gives,
\begin{align}\label{5XII24.f1}
N^i&=\bar{W}\frac{2\bm\br}{\barUh}\frac{\baa_i}{\br^2+\baa_i^2}
\Bigg(
1-\frac{2\bm\br}{\barUh}\frac{\sum_{j=1}^N\frac{a_j^2\red{(\bmu^j)^2}}{\bXi_j(\br^2+\baa_j^2)}}{1+\frac{2\bm\br}{\barUh}u^TA^{-1}u }
\Bigg)\nonumber\\
&=\left(\frac{\bW{2\bm\br}}{{\barUh}+{2\bm\br}\,u^TA^{-1}u }\right) \frac{\baa_i}{\br^2+\baa_i^2}\,,
\end{align}
where again $u^T A^{-1}u=\sum_{i=1}^N\frac{\baa_i^2\bmuisq}{(\br^2+\baa_i^2)\bXi_i}$.
Therefore the norm is,
\begin{align}
N_ih^{ij}N_j
&=\left(\bW\frac{2\bm\br}{\barUh}\right)^2\,\frac{u^TA^{-1}u}{1+\frac{2\bm\br}{\barUh}\,u^TA^{-1}u}\,.
\end{align}
Now we can obtain $\pm N^2$ from \eqref{20XI24.f4};
\begin{align}
\pm N^2&= \bar{W}(1-\ellsquare\bar{r}^2) 
+ \frac{2\bar{m}\br}{\barUh}\bar{W}^2 \left(1 -\frac{2\bm \br}{\barUh}\frac{u^TA^{-1}u}{1+\frac{2\bm\br}{\barUh}\,u^TA^{-1}u}\right)
\nonumber\\
&= \frac{\bar{W}}{1+\frac{2\bm\br}{\barUh}\,u^TA^{-1}u}\left((1-\ellsquare\bar{r}^2)\left(1+\frac{2\bm\br}{\barUh}\,u^TA^{-1}u\right) +\bW \frac{2\bar{m}\br}{\barUh}\right)\,.
\end{align}
This can be further simplified by noting
\begin{equation}
(1-\ell^{-2}\br^2)u^TA^{-1}u+\bW=1+\sum_{i=1}^{N}\left(\frac{(1-\ellsquare\br^2)}{\br^2+\baa_i^2}+\frac{1}{\ell^2}\right)\frac{\baa_i^2\bmuisq}{\bXi_i}=1+\sum_{i=1}^{N}\frac{\baa_i^2\bmuisq}{\br^2+\baa_i^2}\,.
\end{equation}
Hence
\begin{align}\label{20XI24.f5}
\pm N^2&= \frac{\bar{W}}{1+\frac{2\bm\br}{\barUh}\,u^TA^{-1}u}\left((1-\ellsquare\bar{r}^2)+\frac{2\bm \br}{\barUh}\left[1+ \sum_{i=1}^{N}\frac{\baa_i^2\bmuisq}{\br^2+\baa_i^2}\right]\right)
\nonumber\\
&=-(\barVh-2\bm\br)\frac{\bW}{\left(1+\frac{2\bm\br}{\barUh}\sum_{i=1}^N\frac{\baa_i^2\bmuisq}{(\br^2+\baa_i^2)\bXi_i}\right)\prod_{i=1}^{N}(\br^2+\baa_i^2)}\,.
\end{align}
Putting everything together we obtain the equivalent of \cite[Equation~(37)]{Gibbons:2004uw}, 
\begin{equation}\label{25VII24f.8ny}
\det \hat{g}=-\bW(\barVh-2\bm\br) \prod_{i=1}^{N}\frac{\bmuisq}{\bXi_i}\,.
\end{equation}
Note that we have recovered \eqref{25VII24f.8} without dividing by $1-\ellsquare \br^2$.  
However, in this alternative calculation we have instead a division-with-cancellation by 
$\left(1+\frac{2\bm\br}{\barUh}\,u^TA^{-1}u\right)$, which vanishes at the boundary of the causality-violating regions. 
Since the previous calculation did not necessitate such a division, and  since the zero-sets do not overlap in the fully-rotating case by \eqref{28VII24.11a}, 
we conclude that  \eqref{25VII24f.8ny} holds without the restrictions mentioned.
Hence:
\begin{eqnarray} 
&& \mbox{
	both the first and second multiplicative factors in \eqref{25VII24f.6}  have constant signs
}
\nonumber
\\
&&
\mbox{ 
	wherever   $\barVh-2\bar{m}\br$ has constant sign. }
\label{6XI24.2}
\end{eqnarray}

We should keep in mind that the metric $\barg $  is defined as the metric induced by \eqref{3VI24.f7} on the constraint set \eqref{3VI24.f5}. 
Nevertheless, for the $\red{\bmu^i}$-sector,  it appears to be convenient to  first treat the variable $\red{\bmu^{N+1}}$ as independent and to impose the constraint later. That is to say, we calculate the determinant of the $\red{\bmu^i}$-sector as follows:
\begin{enumerate}
	\item We write  $x^I=(\red{\bmu^i}, \bar \mu_{N+1})$, $1\le i \le N$, and ignore that these variables  are constrained.
	We calculate $\det(\tilde g_{IJ})$.
	
	\item Replace the coordinate $\bar\mu_{N+1}$ by a new coordinate $\Phi$ defined as
	\begin{equation}\label{25VII24f.1}
	\Phi 
	:= \bmu^2_{N+1} - \sum_{i=1}^{N} \bmuisq -1\,,
	\end{equation}
	and rewrite the metric in these new coordinates $x^{\bar{I}}:=(x^i,\Phi):=(\red{\bmu^i},\Phi)$, where $1\leq i\leq N$. The determinant we are interested in is the determinant of the induced metric, say $\tilde \gamma_{ij}$, on the hypersurface $\Phi=0$.
	
	\item In these new coordinates we have the following relation
	\begin{equation}
	\det(\tilde g_{IJ})=4\mu_{N+1}^2\det(\tilde g_{\bar{I}\bar{J}})\,,
	\end{equation}
	which follows from the Jacobian of this coordinate transformation.
	
	\item 
	
	We do an ADM-like decomposition of the metric $\tilde g_{\bar{I}\bar{J}}$.
	\begin{equation}
	\tilde g_{\bar{I}\bar{J}}dx^{\bar{I}}dx^{\bar{J}}={\cal N}^2d\Phi^2+\tilde \gamma_{ij}(dx^{i} +{\cal N}^{i}d\Phi)(dx^{j} +{\cal N}^{j}d\Phi)\,.
	\end{equation}
	In this notation
	\begin{equation}
	\det(	\tilde g_{\bar{I}\bar{J}} )={\cal N}^2 \det(\tilde \gamma_{ij})\,.
	\end{equation}
	Note that we also need  to make sure that a zero or infinity of $\cal N$ is \emph{not} compensated by an infinity or zero of $\det(\tilde \gamma_{ij})$; this will be addressed in due course. 
	
	\item Finally we calculate $\cal N$ from the expression ${\cal N}^{-2}= (\partial_{ \bar{I} }\Phi) \tilde g^{\bar{I}\bar{J}} (\partial_{\bar{J}}\Phi)$ and use this to get the desired determinant 
	\begin{equation}\label{25VIIf.3}
	\det(\tilde \gamma_{ij})=\frac{1}{4\mu_{N+1}^2 {\cal N}^2}\det(\tilde g_{IJ})\Big|_{\Phi=0}\,.
	\end{equation}
\end{enumerate}

We begin with step 1. 
For $\br^2 \ne \ell^2$
the $\red{\bmu^i}$-sector of the metric, i.e.
\begin{align*}
&
\sum_{i=1}^N\frac{\bar{r}^2+\baa_i^2}{\bar{\Xi}_i}
\dbmuisq 
- \bar{r}^2
\red{(d\red{\bmu^{N+1}})^2}
-\frac{\ellsquare }{\bar{W}(1 -\ellsquare \bar{r}^2)}\Bigl(\sum_{i=1}^N \frac{\bar{r}^2+\baa_i^2}{\bar{\Xi}_i}\red{\bmu^i} d\red{\bmu^i}
- \bar{r}^2\red{\bmu^{N+1}}d\red{\bmu^{N+1}}\Bigr)^2
\,,
\end{align*}
is in the appropriate form to apply the matrix determinant lemma with $M=D+uu^T$, where 
\begin{align}
uu^T&=-\frac{\palphasquare }{\bar{W}(1 -\ellsquare \bar{r}^2)}
\Bigl(\sum_{i=1}^{N} \frac{\bar{r}^2+\baa_i^2}{\bar{\Xi}_i} \red{\bmu^i} d\red{\bmu^i}
- \bar{r}^2  \red{\bmu^{N+1}}d\red{\bmu^{N+1}}\Bigr)^2\,,
\\
D&=\sum_{i=1}^{N}\frac{\bar{r}^2+\baa_i^2}{\bar{\Xi}_i}\dbmuisq - \bar{r}^2 \red{(d\red{\bmu^{N+1}})^2}\,.
\end{align}
This gives
\begin{align}\label{25VIIf.5}
\det(\tilde g_{IJ})&=-\left(1-\frac{\ellsquare }{\bar{W}(1 -\ellsquare \bar{r}^2)}
\left[
\sum_{i=1}^{N}
\frac{\bar{r}^2+\baa_i^2}{\bar{\Xi}_i} \bmuisq
- \bar{r}^2  \red{(\bmu^{N+1})^2} \right]
\right)
\prod_{i=1}^{N+1} \frac{\bar{r}^2+\baa_i^2}{\bar{\Xi}_i}
\nonumber\\
&=
-\frac{1}
{\bar{W}(1 -\ellsquare \bar{r}^2)}
\underbrace{\left(-\sum_{i=1}^{N}\bmuisq+\red{(\bmu^{N+1})^2}\right)}_{= \Phi+1}
\bar{r}^{2}
\prod_{i=1}^{N} \frac{\bar{r}^2+\baa_i^2}{\bar{\Xi}_i}\,.
\end{align}

Steps 2-4 are taken care of since they only set up the problem. In particular, since we did not change the coordinates $\red{\bmu^i}$ for $1\leq i\leq N$ we
have simply $\tilde \gamma_{i j}=\bar g_{i j}$. Thus, the only thing left to calculate is
\begin{equation}\label{25VIIf.2}
{\cal N}^{-2}= (\partial_{ \bar{I} }\Phi) \tilde g^{\bar{I}\bar{J}} (\partial_{\bar{J}}\Phi)\,.
\end{equation}
Since this is a scalar we can work in the original coordinate system and calculate the gradient of $\Phi$ given by \eqref{25VII24f.1}. This yields,
\begin{equation}
d\Phi=2\left(\sum_{i=1}^{N}\red{\bmu^i}d\red{\bmu^i} -\red{\bmu^{N+1}} d\red{\bmu^{N+1}}\right)\,.
\end{equation}

To calculate the inverse metric we plug \eqref{24VIIf.3} into \eqref{25VIIf.2} to obtain
(note that the expression $1+u\cdot D^{-1}\cdot u $ appearing below has already been calculated in \eqref{25VIIf.5})
\begin{align}\label{25VIIf.4}
{\cal N}^{-2}&=d\Phi\cdot D^{-1}\cdot d\Phi -\frac{(u\cdot D^{-1}\cdot d\Phi)^2}{1+u\cdot D^{-1}\cdot u}
\nonumber\\
&= - 4 \left(\frac{\red{(\bmu^{N+1})^2}}{\bar{r}^2 } - \sum_{i=1}^{N}\frac{\bmuisq\,\bar{\Xi}_{i}}{\bar{r}^2+\baa_i^2}\right)
\nonumber\\
&\quad
-\frac{4\bar{W}(1 -\ellsquare \bar{r}^2)}{\left(\sum_{i=1}^{N}\bmuisq - \red{(\bmu^{N+1})^2}\right)}\frac{\palphasquare }{\bar{W}(1 -\ellsquare \bar{r}^2)}
\left(\red{(\bmu^{N+1})^2} - \sum_{i=1}^{N}\bmuisq \right)^2
\nonumber\\
&= - 4(1-\ellsquare \bar{r}^2)\left(\frac{\red{(\bmu^{N+1})^2}}{\bar{r}^2 } - \sum_{i=1}^{N}\frac{\bmuisq}{\bar{r}^2+\baa_i^2}\right
)
\nonumber\\
& 
= - \frac{4(1-\ellsquare \bar{r}^2)}{\bar r^2}
\left(
1  + \sum_{i=1}^{N}\frac{\baa_i^2 \bmuisq}{\bar{r}^2+\baa_i^2}
\right ) 
\,.
\end{align}

All that is left is to put the pieces together. Substituting, \eqref{25VIIf.5} and \eqref{25VIIf.4} into \eqref{25VIIf.3} we obtain
\begin{equation}\label{25VII24f.7}
\det(\tilde g_{ij}) = \det(\tilde \gamma_{ij})=
\frac{1}{\bar{W}\,\red{(\bmu^{N+1})^2}}
\left(
1  + \sum_{i=1}^{N}\frac{\baa_i^2 \bmuisq}{\bar{r}^2+\baa_i^2}
\right )
\prod_{i=1}^{N} \frac{\bar{r}^2+\baa_i^2}{\bar{\Xi}_i}\,.
\end{equation}

While this formula strongly suggests that the tensor field $\tilde g _{ij}$ is non-degenerate everywhere, there is a catch in that the calculation  involves a cancellation of a multiplication and a division by $\br^2-\ell^2$. 
Indeed, the expression for ${\cal N}^{-2}$ has a zero at $\br = \pm \ell$  which is compensated by a factor in \eqref{25VIIf.5} in the final formula for the determinant of $\tilde{g}_{ij}$.
So an alternative calculation of the  determinant at  $\br = \pm \ell$ is needed.
For this we can use directly the form of the metric in \eqref{3VI24.f78}, since at $\br =\pm\ell$ the $\red{\bmu^i}$-sector of the metric (with the constraint now imposed) is in a form to which \eqref{24VIIf.2} applies.
In particular, when $\br = \pm \ell$  the prefactor in front of the $(d\bar{W})^2$ terms vanishes and we have
\begin{align}
\tilde{g}_{ij}   \Big|_{\br^2=\ell^2}&=  \sum_{i=1}^N\frac{\ell^2+\baa_i^2}{\bar{\Xi}_i}\dbmuisq  
- \ell^2 
\frac {\Big( \tfrac{1}{2}\sum_{i=1}^{N} d(\bmuisq )\Big)^2}{1 +   \sum_{i=1}^{N} \bmuisq}
\nonumber\\
&=\ell^2\left(\sum_{i=1}^N\dbmuisq  -\frac{\left(\sum_{i=1}^N\red{\bmu^i}d\red{\bmu^i}\right)^2}{1 +   \sum_{i=1}^{N} \bmuisq}\right)\,.
\label{12XI24.51}
\end{align}
Thus,
\begin{align}
\det\left(\tilde{g}_{ij}   \Big|_{\br^2=\ell^2}\right)&=\ell^{2N}\left(1-\frac{\sum_{i=1}^N\bmuisq}{1 +   \sum_{i=1}^{N} \bmuisq}\right)=\frac{\ell^{2N}}{\red{(\bmu^{N+1})^2}}\,,
\end{align}
which, by the way, coincides with \eqref{25VII24f.7} evaluated at $\br^2=\ell^2$. The last equation shows that the zero of ${\cal N}^{-2}$ is innocuous, and the determinant  of the $\red{\bmu^i}$-sector of $\tilde g$ has no zeros.

Note that  the   $\red{\bmu^i}$-part of $\tilde g$ in \eqref{12XI24.51} is Riemannian when all the $\red{\bmu^i}$'s vanish, and by continuity the signature of this part of $\bar g $ is Riemannian everywhere.

Now, going back to \eqref{25VII24f.6} and substituting \eqref{25VII24f.7}, \eqref{25VII24f.8}, and for $g_{\bar{r}\bar{r}}$, the determinant of the full metric reads 
\begin{align}\label{26VII24f.1}
\det (\bar g_{\mu\nu})
&=\bar g_{\br\br} \det(\hat g_{\alpha\beta})\det(\tilde g_{i j})
\nonumber
\\
&=-
\frac{\barUh}{\barVh-2\bm\br}
\bar{W}
\left(\barVh-2\bar{m}\bar{r}\right)
\frac{1}{\bar{W}\,\red{(\bmu^{N+1})^2}}
\left(
1  + \sum_{i=1}^{N}\frac{\baa_i^2 \bmuisq}{\bar{r}^2+\baa_i^2}
\right )\,\prod_{i=1}^{N} \frac{\bar{r}^2+\baa_i^2}{\bar{\Xi}_i}\prod_{i=1}^{N}\frac{\bmuisq}{\bar{\Xi}_i}
\nonumber\\
& 
=-
\frac{1}{\red{(\bmu^{N+1})^2}}
\left(1 +   \sum_{i=1}^{N}\frac{\baa_i \bmuisq}{\bar{r}^2+\baa_i^2}
\right)^2\prod_{i=1}^{N} \frac{(\bar{r}^2+\baa_i^2)^2\,\bmuisq}{\bar{\Xi}_i^2}
\,.
\end{align}
This finishes the proof of Proposition~\ref{P6XI24.1}.

\section{Zeros of $\barVh - 2 \br \bar m$} \label{app9XI24.1}

In order to simplify notation we rescale $\br$, $\bm$, and the rotation parameters $\baa_i$ so that $\ell=1$: 
\begin{equation}\label{18XI24.1}
x:= \frac{\br}{\ell}
\,,
\quad
a_i:= \frac{\baa_i}{\ell}
\,,
\quad
m:=  \frac{\bm}{\ell^{2N-1}}
\,.
\end{equation}
The zeros of $\barVh - 2 \br \bar m$ are then roots of the polynomial
\begin{align}\label{9XI24.111} 
\hat{F}(x):= (1-x^2)
\prod_{i=1}^\red{N}(x^2+\nobara_i^2)
+
2 m x
\,.
\end{align}
We will show that $\hat F(x)$ has at most four real zeros. The case $N=1$ is obvious, as $\hat F(x)$ is a polynomial of degree four, hence with at most four roots. Thus we assume $N\geq 2$.

First, we expand the polynomial $\hat F(x)$ by noting that
%the product in \eqref{9XI24.11} is a symmetric polynomial which can be written
%
\begin{equation}
\prod_{i=1}^\red{N}(x^2+\nobara_i^2)=\sum_{k=0}^{N}A_k x^{2(N-k)}
\,,\quad \text{where for $k\ge 2$,}\quad
A_{k}:=\sum_{1\le i_1<\ldots < i_k\le N}  a_{i_1}^{2}\cdots a_{i_k}^2\,,
\end{equation}
are the elementary symmetric polynomials with
$$
A_{0}:=1
% \,,
%  \quad
%  A_1 :=\sum_{  i =1} ^Na_i^2 
\,.
$$
It follows readily that
%.
%%
\begin{equation}
\hat F(x)=-x^{2N +2 } -\sum_{k=0}^{N-1}
(A_{k+1}-A_{k})x^{2(N-k)} +
2m x+A_N\,,
\end{equation}
where the coefficients of the powers of $x$ can be written as
\begin{align}
%h_1 
A_1-A_{0}
&=  -1 + \sum_{i=1}^{N} a_i^2   \,,
\\
A_2-A_{1}
% h_2 
& = \sum_{i=1}^{N} a_i^2 \Big(-1 + \sum_{k>i}^{N} a_k^2\Big)
\,,
\\
A_3-A_{2}
%     h_3 
&= \sum_{i=1}^{N-1} a_i^2 \Big[\sum_{k>i}^{N} a_k^2 \Big(-1 + \sum_{j>k}^{N} a_{j}^2 \Big)\Big]
\,,
\\
&\vdots
\\
A_{k}-A_{k-1}
& = \sum_{i_1=1}^{N+2-k} \Big\{ a_{i_1}^2
\Big[ \sum_{i_2 > i_1}^{N+3-k} a_{i_2}^2 
\Big(
\sum_{i_3>i_2}^{N+4-k}a_{i_3}^2
\cdots 
\sum_{i_{k-1}>i_{k-2}}^{N-1} a_{i_{k-1}}^2
\big(-1 + \sum_{i_k> i_{k-1}}^{N} a_{i_k}^2
\big)
\Big)
\Big]
\Big\}\,.
\label{21XI24.1}
\end{align}
To better reflect the recursive nature of these coefficients, it is convenient to introduce the following definitions:
(here and below, if the upper limit of a sum is smaller than the lower limit, then the sum is 
taken to be zero)%  %
\begin{align}
\label{21XI24.w1}
h_1^{[k]}&:= -1 + \sum_{i > k}^N a_i^2
\,,\quad
h_\ell^{[k]} : = \sum_{i > k}^{N+2-\ell} a_{i}^2 h_{\ell-1}^{[i]}
%    \nn
%    \\
%    & = \sum_{i_1 > k}^{N+2-\ell} \Big\{ a_{i_1}^2
%    \Big[ \sum_{i_2 > i_1}^{N+3-\ell} a_{i_2}^2 
%    \Big(
%    \sum_{i_3>i_2}^{N+4-\ell}
% \cdots \big(-1 + \sum_{i_k> i_{k-1}}^{N} a_{i_k}^2
% \big)
% \Big)
% \Big]
% \Big\} 
\,, \qquad
\ell \in \N\cap[2,N+1]\,, k \in \Z\cap[0,N]
\,.
\end{align}
Writing-out the last definition in full:
\begin{align}
\label{21XI24.w1b} 
h_{\ell}^{[k]} 
& = \sum_{i_1 > k}^{N+2-\ell} \Big\{ a_{i_1}^2
\underbrace{\Big[ \sum_{i_2 > i_1}^{N+3-\ell} a_{i_2}^2 
	\underbrace{\Big(
		\sum_{i_3>i_2}^{N+4-\ell}
		\cdots \underbrace{\big(-1 + \sum_{i_\ell> i_{\ell-1}}^{N} a_{i_\ell}^2
			\big)}_{h_{1}^{[i_\ell]}}
		\Big)}_{h_{\ell-2}^{[i_2]}}
	\Big]
	\Big\} }_{h_{\ell-1}^{[i_1]}} 
% \,,
%    \\
%    h_{N+1}^{[0]} &= -a_1^2a_2^2\dots a_N^2 
\,.
\end{align}
Comparing \eqref{21XI24.1} and \eqref{21XI24.w1b} one finds
\begin{align}
\label{21XI24.w1bc}
% h_1^{[0]} &= -1 + a_1^2+a_2^2+\dots+a_N^2 \,,
% \quad 
h_\ell^{[0]} &= A_{\ell}-A_{\ell-1} 
\,,
\end{align}
hence 
\begin{align}
\hat F(x)= -x^{2(N+1)}- \sum_{i=1}^{N} h_i^{[0]} x^{2(N-(i-1))} + 2mx + a_1^2a_2^2\dots a_N^2\,.
\end{align}

To see that $\hat F(x)$ has  at most  four real roots we make use of Descartes' rule of signs, which states that the number of positive real roots of a polynomial is not larger than the number of sign changes in the sequence of polynomial's non-zero coefficients, while the number of real negative roots is no larger than the number of sign changes after multiplying the coefficients of odd-power terms by $-1$.  
In our case, we are done if we can show:

\begin{lemma}
	\label{L21XI24.1}
	There can be at most one change of signs in the sequence $(1, h_1^{[0]},h_2^{[0]},...,h_N^{[0]})$
\end{lemma}

{\sc\noindent Proof}: If all $h_i^{[0]}$'s are positive then there are no changes of signs and we are done. We will prove the lemma by showing that for integers $i\in [1,N]$ and $k_0\in [0,N]$ we have the following implication:
\begin{align}
h_i^{[k_0]} \leq 0 \implies h_i^{[k]} \leq 0 \quad \forall \ k\in \Z\cap[k_0,N]\,.
\label{22XI24.1}
\end{align}
Assuming that \eqref{22XI24.1} is true,  setting $k_0=0$ there gives 
$$
h_i^{[0]} \leq  0 \implies h_i^{[k\geq 0]} \leq 0 \implies h_{i+1}^{[0]}\leq 0
\,,
$$
where the last implication follows from the fact that $h_{i+1}^{[0]}$ is a sum of terms of the form $a_k^2 h_{i}^{[k]}$ (cf. \eqref{21XI24.w1}). Thus there can only be one change of signs. 

Given  $0 \leq k_0 \leq N$, we will prove the implication \eqref{22XI24.1} by induction on $i$.

\paragraph{$i=1$:} We have 
\begin{align}
h_1^{[k]} = -1 + a_{k+1}^2 + a_{k+2}^2 + ... + a_{N}^2 \,. 
\end{align}
Clearly,
$$
h_1^{[k+1 ]} = -1 + a_{k+2}^2 + \ldots  + a_{N}^2 <  -1 + a_{k+1}^2 + a_{k+2}^2 +\ldots  + a_{N}^2 =  h_1^{[k ]} 
\,.
$$
Indeed,  $h_1^{[k]}  < h_1^{[k_0]}  $ for all $k \geq k_0$,  so if $h_1^{[k_0]} \leq 0$ then $h_1^{[k]} \leq 0$
for all
$k \geq k_0$.

\paragraph{$i=j_0+1$:} 
Assuming \eqref{22XI24.1} is true for $i=j_0$, we now show that it is true for $i=j_0+1$. 
Thus, we assume $h_{j_0+1}^{[k_0]} \leq 0$:
\begin{align}
h_{j_0+1}^{[k_0]} &= a_{k_0+1}^2 h_{j_0}^{[k_0+1]} + a_{k_0+2}^2 h_{j_0}^{[k_0+2]}
% +...+ a_{\ell_0}^2h_{j_0}^{[\ell_0]} 
% + a_{\ell_0+1}^2h_{j_0}^{[\ell_0+1]} 
+ ...
+ a_{N+1-j_0}^2h_{j_0}^{[N+1-j_0]}
\leq 0  \,.
\label{22XI24.2a}
\end{align}
If all the terms in the sum are negative there is nothing to prove. Thus, suppose not. Then because of the $i=j_0$ statement, we can assume that there is some $k_0 + 1< \ell_0 \leq N$ such that $h_{j_0}^{[i]} \geq 0$ for $k_0 \leq i \leq \ell_0$ and $h_{j_0}^{[i]} < 0$ for $\ell_0 < i \leq N$, i.e.,
\begin{align}
h_{j_0+1}^{[k_0]} &= \underbrace{a_{k_0+1}^2 h_{j_0}^{[k_0+1]} + a_{k_0+2}^2 h_{j_0}^{[k_0+2]}+...+ a_{\ell_0}^2h_{j_0}^{[\ell_0]}}_{\text{each term is } \geq 0} 
+ \underbrace{a_{\ell_0+1}^2h_{j_0}^{[\ell_0+1]} + ...+ a_{N+1-j_0}^2h_{j_0}^{[N+1-j_0]}}_{\text{each term is }< 0}
\nn
\\
&
\leq 0  \,.
\label{22XI24.2b}
\end{align}
If we start removing terms from $h_{j_0+1}^{[k_0]}$ in \eqref{22XI24.2b} starting from terms from the left, clearly the left-over sum will always remain negative. 
Since the $h_{j_0+1}^{[k\geq k_0]}$'s are exactly obtained in this way, they are all negative, and thus the claim for $i=j_0+1$ is true.  
\qed 
\section{Regularity of $\myH_\pm$}
\label{s17XI24.1}

In this appendix we show that the  functions $\myH_\pm$  of \eqref{26XI24.1}, i.e. 
\begin{equation}\label{26XI24.1b}
\myH_-(\br):=\inf_{\red{\bmu^i}\in \R^N } \frac{ \breve U^{-1}N ^2 }{
	|\barVh-2\bar{m}\br|}
\ \mbox{and} 
\
\myH_+(\br):= \sup_{\red{\bmu^i}\in \R^N } \  \frac{ \breve U^{-1}N ^2 }{|\barVh-2\bar{m}\br|}
\,,
\end{equation}
are Lipschitz-continuous. 
This follows from Lemma~\ref{L30XI24.1} below after some preparatory work. We expect these functions to be piecewise smooth, but whether or not they are is irrelevant for the applications in this work.
We consider a fully-rotating configuration.  

Recall that 
\begin{align}\label{26XI24.6} 
\bUh  
= \ & 
\underbrace{\bigg(1
	+ \sum_{i=1}^N \frac{ {\bmuisq}\baa_i^2}
	{\br^2+\baa_i^2}\bigg)
}_{
	\brU } \prod_{i=1}^N (\bar{r}^2+\baa_i^2) =:
\brU  \myF ^{2}
\,,
\\
N^2 = \ & 
\frac{\bW |\barVh-2\bm\br|}{\left(1+\frac{2\bm\br}{\barUh}
	\sum_{i=1}^N\frac{\baa_i^2\bmuisq}{(\br^2+\baa_i^2)\bXi_i}\right)\prod_{i=1}^{N}(\br^2+\baa_i^2)}
\nonumber
\\ = \ & 
\frac{\brU   \bW |\barVh-2\bm\br|}
{ \brU  \myF ^{2}+ {2\bm\br} 
	\sum_{i=1}^N\frac{\baa_i^2\bmuisq}{(\br^2+\baa_i^2)\bXi_i} 
}  
\,,
\\	
\bar W
= \ & 
1 +  \sum_{i=1}^{N }
\frac{ \bar a_i^2\bmuisq}{\ell^2 \bar\Xi_{i}}
\,,
\\	
\myPhi: = \ & 
\frac{ \breve U^{-1}N ^2 }{
	|\barVh-2\bar{m}\br|}  
\nonumber
%    \\	 
%      = \ & 
%     \frac{ 
%      1 +  \sum_{i=1}^{N }
%		\frac{ \bar a_i^2\bmuisq}{\ell^2 \bar\Xi_{i}}
%         }
%          {\brU  \left(1+\frac{2\bm\br}{\barUh}
%            \sum_{i=1}^N\frac{\baa_i^2\bmuisq}{(\br^2+\baa_i^2)\bXi_i}\right)\prod_{i=1}^{N}(\br^2+\baa_i^2)
%                }
%  \nonumber
\\ = \ &    \frac{ 
	1 +  \sum_{i=1}^{N }
	\frac{ \bar a_i^2\bmuisq}{\ell^2 \bar\Xi_{i}}
}
{
	\big(1
	+ \sum_{i=1}^N \frac{ {\bmuisq}\baa_i^2}
	{\br^2+\baa_i^2}\big)
	\prod_{i=1}^N (\bar{r}^2+\baa_i^2) 
	+ {2\bm\br} 
	\sum_{i=1}^N\frac{\baa_i^2\bmuisq}{(\br^2+\baa_i^2)\bXi_i} 
}  
\,.
\end{align} 

We note that the function $\myPhi$ has an obviously vanishing derivative at $\red{\bmu^i}=0$, with 
\begin{equation}\label{26XI24.7}
\myPhi|_{\red{\bmu^i}=0} =   \prod_{i=1}^N (\bar{r}^2+\baa_i^2) ^{-1}
\,.
\end{equation} 
If it were a global extremum for some values of parameters, smoothness either of the $\sup$ or  of the $\inf$, whichever relevant, would immediately follow.
We have not pursued this question any further.  

To continue, it is useful to analyse the asymptotic behaviour of $ \breve U^{-1}N^2$. For this we  set
\begin{equation}\label{16XI24.11a}
|\bmu| = \sqrt{\sum \bmuisq}
\,,
\quad
\blue{n^i}:= |\bmu|^{-1} \red{\bmu^i}
\,.
\end{equation}
We have 
\begin{align}\label{16XI24.8} 
\lim_{|\bmu| \to\infty}   	\myPhi \equiv  \ & 
\lim_{|\bmu| \to\infty}  \frac{ \breve U^{-1}N ^2 }{
	|\barVh-2\bar{m}\br|}  
\nonumber
\\	  
= \ &   \frac{    \sum_{i=1}^{N }
	\frac{ \bar a_i^2 \red{(n^i)^2}}{\ell^2 \bar\Xi_{i}}
}
{
	\big( \sum_{i=1}^N \frac{ \red{(n^i)^2}\baa_i^2}
	{\br^2+\baa_i^2}\big)
	\prod_{i=1}^N (\bar{r}^2+\baa_i^2) 
	+ {2\bm\br} 
	\sum_{i=1}^N\frac{\baa_i^2\red{(n^i)^2}}{(\br^2+\baa_i^2)\bXi_i} 
}  
\nonumber
\\
=:  \ &  
\myHS
\,.  
\end{align}
The function $\myHS$  is a rational function of expressions of the form 
$$
\sum \alpha_i^2 \baa_i^2 \red{(n^i)^2} = \vec X \cdot \vec N
\,,
$$
where $\cdot$ denotes the  Euclidean scalar product, with
$$
\vec X:=(\alpha_1^2 \baa_1^2,\ldots, \alpha_N^2 \baa_N^2) 
\ \mbox{and}
\ 
\vec N:=
(n_1^2, \ldots, n_N^2)
\,,
$$
%, 
for some coefficients $\alpha_i>0$.
Thus both $\vec X$ and $\vec N$ lie in the positive quadrant of $\R^N$. 
Since all the  $\alpha_i^2 a_i^2$'s are non-vanishing, $\vec X $ and $\vec N$ are never orthogonal. 
It follows that the angle between $\vec X$ and $\vec N$ is in $[0,\pi/2)$. Since  $\vec X \cdot \vec N = |\vec X| |\vec N|  \cos (\theta)$ we see that each  such scalar product occurring in $\myHS$ 
is strictly positive. 
We conclude that the function $\myHS(\br,\cdot)$ is bounded both from above and from below for each $\br$ on the set defined by the inequality 
\begin{equation}\label{27XI24.1}
\inf_{\blue{n^i}\in S^{n-1}}  
\left[ 
\big( \sum_{i=1}^N \frac{ \red{(n^i)^2}\baa_i^2}
{\br^2+\baa_i^2}\big)
\prod_{i=1}^N (\bar{r}^2+\baa_i^2) 
+ {2\bm\br} 
\sum_{i=1}^N\frac{\baa_i^2\red{(n^i)^2}}{(\br^2+\baa_i^2)\bXi_i} 
\right]
> 0 
\,.
\end{equation}
This further implies that $\myH(\br,\cdot)$ is likewise bounded from above and below on this same set. 

It should be clear, e.g.\ by comparing \eqref{25XI24.21} with the calculations above (see also Proposition~\ref{P30XI24.1}), that the complement of the set defined by \eqref{27XI24.1} coincides with the projection of the closure of the time-machine set, as defined at the end of Section~\ref{sss11XI24.1}. 

Let
$$\Psi:B^N\mapsto  \R^N
$$
be any diffeomorphism from an open ball $B^N$ to $\R^n$ chosen so that for each $\br$ the function 
$$
\myH(\br.\cdot):= H\circ\Psi(\br,\cdot) : B^N\mapsto\R
$$
extends to a smooth function, still denoted by $\myH(\br,\cdot)$, on $\bB^N$. 
For example, 
$$
\red{\bmu^i}= \Psi^i(\vec x) :=  \frac{\red{x^i}}{1-|\vec x|^2}
\,,
$$
but any such map will do.
Note that with this last choice the restriction of $\myH$ to $S^{N-1}$ coincides directly with the function $ \myHS$ already defined in \eqref{16XI24.8}. 

Local Lipschitz-continuity of the functions $\myH_\pm$ follows now from the following standard fact, which we prove for completeness:

\begin{Lemma}
	\label{L30XI24.1}
	Let $\myH$ be a differentiable function on  $I\times \bB^N$, where $I$ is a closed interval and   $\bB^N$ is a compact manifold, possibly with boundary. Then, for $r\in I$ the functions $\inf_{\bB^N} \myH(r,\cdot)$ and $\sup_{\bB^N} \myH(r,\cdot)$ are   Lipschitz functions of $r$.
\end{Lemma}

\proof
For $r$ and $r+\Delta $ in $I$, 
with $\Delta \ge 0$, we can write
\begin{eqnarray}
\myH(r+\Delta,\red{x^i}) 
& = & 
\myH(r, \red{x^i})  + \int_0^1 \frac{d
	\Big( \myH\big(t (r+ \Delta)+(1-t)r,\red{x^i}\big)\Big)
}
{dt} 
\, dt
\nonumber
\\   
& = & 
\myH(r,  \red{x^i})  + \Delta  \int_0^1 (\partial_r 
\Big( \myH\big(t (r+ \Delta)+(1-t)r,\red{x^i}\big)\Big)
\, dt
\,.
\label{30XI24.21}
\end{eqnarray}
Hence
\begin{eqnarray}
\sup_{\red{x^i} \in \bB^N}  \myH(r+\Delta,\red{x^i})  
& = & 
\sup_{\red{x^i} \in \bB^N} 
\Big[\myH(r,\red{x^i})  + \Delta  \int_0^1 (\partial_r 
\Big( \myH\big(t (r+ \Delta)+(1-t)r,\red{x^i}\big)\Big)
\, dt
\Big]
\nonumber
\\    
& \le & 
\sup_{\red{x^i} \in \bB^N} 
\myH(r,\red{x^i})  + 
\Delta  \int_0^1
\sup_{\red{x^i} \in \bB^N,\, r\in I}  (\partial_r 
\Big( \myH\big(t (r+ \Delta)+(1-t)r,\red{x^i}\big)\Big)
\, dt
\nonumber
\\    
&= & 
\sup_{\red{x^i} \in \bB^N} 
\myH(r,\red{x^i})  + 
\Delta   
\sup_{\red{x^i} \in \bB^N, \,r\in I}  (\partial_r 
\myH)
\big(r,\red{x^i}\big) 
\,.
\label{30XI24.2}
\end{eqnarray}
Next, consider \eqref{30XI24.21} at a point $\red{x^i}$ where the sup of  
$\myH(r ,\cdot) $
is attained: 
\begin{eqnarray}
\myH(r+\Delta,\red{x^i})  
& =&     
\sup_{\red{x^i} \in \bB^N} 
\myH(r,\red{x^i})   + \Delta  \int_0^1 (\partial_r 
\Big( \myH\big(t (r+ \Delta)+(1-t)r,\red{x^i}\big)\Big)
\, dt 
\,.
\label{30XI24.4}
\end{eqnarray}
Hence 
\begin{eqnarray}
\sup_ { \bB^N}  \myH(r+\Delta,\cdot)  
& \ge &     
\myH(r+\Delta,\red{x^i})  
\nonumber
\\
& = & 
\sup_{\red{x^i} \in \bB^N} 
\myH(r,\red{x^i})   + \Delta  \int_0^1 (\partial_r 
\Big( \myH\big(t (r+ \Delta)+(1-t)r,\red{x^i}\big)\Big)
\, dt
\nonumber
\\    
& \ge & 
\sup_{\red{x^i} \in \bB^N} 
\myH(r,\red{x^i}) - 
\Delta  \int_0^1
\sup_{\red{x^i} \in \bB^N,\,  r\in I} 
\left| (\partial_r 
\Big( \myH\big(t (r+ \Delta)+(1-t)r,\red{x^i}\big)\Big)
\right|
\, dt
\nonumber
\\    
& =  & 
\sup_{\red{x^i} \in \bB^N} 
\myH(r,\red{x^i})  -
\Delta   
\left|
\sup_{\red{x^i} \in \bB^N, \,r\in I}  (\partial_r 
\myH)
\big(r,\red{x^i}\big) 
\right|
\,,
\label{30XI24.5}
\end{eqnarray}
which finishes the proof of Lipschitz-continuity of $ \sup_{\red{x^i} \in \bB^N} 
\myH(r,\red{x^i})$. 

Since $\inf f = - \sup (-f)$,  applying what has been established so far to $-\myH$ finishes the proof of the Lemma.
\qedskip

{%\blueref{appendix added}
}
\section{Analytic continuation}
\label{s20V25.1} 
\newcommand{\Ric}{\blue{\mathrm{Ric}}}
\newcommand{\Ricci}{\Ric}
\newcommand{\bRic}{\overline{\Ric}}

Consider a manifold $\mcM$ of dimension $d$ equipped with a non-degenerate tensor field $g=g_{\mu\nu} dx^\mu dx^\nu$ satisfying, say, the vacuum Einstein equations with a cosmological constant,
\begin{equation}\label{21V25.1}
\Ric = \lambda g
\,,
\end{equation}
for some $\lambda \in \R$; the signature is irrelevant for what follows. Suppose that all the metric functions $g_{\mu\nu}$ are real analytic in a coordinate $x^1$ near a point  $p\in \mcM$. More precisely, let us assume there exists a neighborhood of $p$ with a coordinate system such that $x^1(p)=0$ and  on which the Taylor series in $x^1$, at $x^1=0$, of each  metric function
$g_{\mu\nu}$ converges for all $|x^1| < R$ to its value $g_{\mu\nu}(x^1,\ldots,x^d)$. 

Consider the open disc $D\subset \C$ of radius $\rho$, and for $z \in D $ define a complex valued tensor field $\bar g_{\mu\nu}(z,x^2,\ldots,x^d)$ by the just-mentioned Taylor series. Restricting $\rho$ if necessary we can without loss of generality assume that $|\det (\bar g)|>0$ on $D$.

We can define the Christoffel symbols, the Riemann tensor, and the  Ricci tensor of $\bar g$ by the usual formulae 
of pseudo-Riemannian geometry, with $\partial_{x^1}$  replaced by $\partial_z $. 
For real values of $z$
we have $\bar g =g$, so that the Ricci tensor of $\bar g$, say $\bRic$, satisfies there
$$
\bRic = \lambda \bar g
\,.
$$
Since $\bRic$ is holomorphic in $z$ this remains true for all $z\in D$. 
For $a\in\C$ define
$$
\psi_a (z ,x^2,\ldots, x^d) =  (a z ,x^2,\ldots, x^d) 
\,.
$$
Set
$$
\bar g_a = \psi_a^* \bar g
\,.
$$
By the usual formulae of tensor calculus, the Ricci tensor $\bRic_a$ of $\bar g_a$
satisfies
\begin{equation}\label{21V25.2}
\bRic_a = \psi_a^* \bRic
\,, 
\end{equation}
%. 
and we conclude that 
\begin{equation}\label{21V25.3}
\bRic_a = \lambda \bar g_a
\,. 
\end{equation}
%. 
Restricting the coordinate $az$ to real-values, and reverting from $\partial_{z}$ to $\partial_{x^1}$, one finds that the Ricci-tensor of the (possibly imaginary-valued) tensor $\bar g_a$ continues to satisfy \eqref{21V25.3}.

Suppose, next, that the metric $\bar g_a$ above depends analytically on a real-valued parameter, say $b$. A similar argument shows that $\bar g_a$ can be analytically extended to all values of $b\in\C$ which belong to a simply connected and  arc-connected Riemann surface on which $|\det g_a|>0$, still satisfying \eqref{21V25.3}.

One can iterate this argument with further coordinates and parameters in the metric, if applicable. 

The task is then to find parameters $a,b\in\C$ as above so that the tensor $\bar g_a$ is real-valued with Lorentzian signature, hence a Lorentzian Einstein metric. In our calculations the coordinates are $t$ and $r$, and the parameter $a$ used is $\sqrt{-1}$   for $t$ and $-\sqrt{-1}$   for $r$. The parameter $b$ here corresponds  to the $a_j$'s and $m$ in the main text.

\bigskip

%\bibliographystyle{JHEP}
%\bibliography{New_Metrics-minimal}

\providecommand{\href}[2]{#2}\begingroup\raggedright\endgroup

\end{document}